\documentclass[12pt]{report}

\newcommand{\erfc}{\mathrm{erfc}}

\newcommand{\beq}{\begin{equation}}
\newcommand{\eeq}{\end{equation}}
\newcommand{\bea}{\begin{eqnarray}}
\newcommand{\eea}{\end{eqnarray}}
\newcommand{\beg}{\begin{align}}
\newcommand{\eeg}{\end{align}}

\def\bary{\begin{array}}
\def\eary{\end{array}}

\def\bmat{\left[ \begin{array}}
\def\emat{\end{array} \right]}

\def\bmatt{\left\{ \begin{array}}
\def\ematt{\end{array} \right.}
\usepackage[latin5]{inputenc}
\usepackage{amssymb,graphicx,amsfonts,amsmath,amssymb,fbe_tez}

\title{IMPULSE RADIO ULTRA-WIDEBAND COMMUNICATION
OVER FREE-SPACE OPTICAL LINKS} \turkcebaslik{OPT\.{I}K
SERBEST-UZAY KANALLARINDA DARBE-RADYO ULTRA-GEN\.{I}\c{S}BANT
HABERLE\c{S}\.{I}M\.{I}} \degree{B.S. Electrical and Electronics
Engineering, Bilkent University, 2008}
\author{Kemal Davasl\i o\u{g}lu}
\program{Electrical and Electronics Engineering}
\supervisor{Assist. Prof. Mutlu Koca} \examineri{Prof. Hakan
Deli{\c c}} \examinerii{Prof. Naci \.{I}nci}
\dateofapproval{29.07.2010}

\begin{document}
\pagenumbering{roman}
%
\makemstitle        

\makeapprovalpage

\begin{acknowledgements}
First of all, I would like to thank Assist. Prof. Mutlu Koca for
his help, encouragement and support throughout my study in
Bo\v{g}azi\c{c}i University. Since the first day I stepped in to
this university, he was there with me. His help has enabled me to
understand and develop a strong intuition on telecommunications
and his way of looking at things has always impressed me. His
critical suggestions have definitely improved my work and this
thesis could not be accomplished without his guidance.

I also wish to thank to the committee members Prof. Hakan Deli{\c
c} and Prof. Naci \.{I}nci for their important suggestions and
valuable comments that have shaped the final version of my thesis.

I am also grateful to Assist. Prof. Ali Emre Pusane and Assoc.
Prof. M. K\i van{\c c} M\i h{\c c}ak for sharing their experience
and valuable ideas on both engineering and career which helped me broaden my approach
and I owe them a lot.

My friends and colleagues were always with me whenever I needed to
chat and with their friendship I was able to clear my mind and
focus clearly. Especially, Erman {\c C}a\u{g}\i ral, Yakup K\i l\i
{\c c}, Berksan \c{S}erbet\c{c}i, O\u{g}uzhan Kondak\c{c}\i, Temu{\c c}in Som, N.
Caner G\"{o}v, Hasan Sicim, Kamil \c{S}enel and Deniz Sevi\c{s}
are just to name a few. Also, I should give a special thanks to
Duygu Eru\c{c}man who has always supported me with her presence
and delicious food.

Last but not least, I am deeply grateful to my parents Ali and
Bilge Davasl\i o\u{g}lu and my grandparents Kemal and G\"{u}lseren
\"{O}zmen for their eternal love and support. I will always have
my utmost and deepest gratitude toward them.

This thesis is supported by the Scientific and Technological
Research Council of Turkey (TUBITAK) under the grant number
105E077.

\end{acknowledgements}

\begin{abstract}
A composite impulse radio ultra-wideband (IR-UWB) communication
system is presented. The proposed system model aims to transmit
UWB pulses over several kilometers through free-space optical
(FSO) links and depending on the link design, the electrical
estimates of the FSO system can be directly used or distributed to
end-user through radio-frequency (RF) links over short ranges.
However, inhomogeneities on the FSO transmission path cause random
fluctuations in the received signal intensity and these effects
induced by atmospheric turbulence closely effect the system
performance. Several distinct probability distributions based on
experimental measurements are used to characterize FSO channels
and using these probabilistic models, detection error probability
analysis of the proposed system for different link designs are
carried out under weak, moderate and strong turbulence conditions.
The results of the analysis show that depending on the atmospheric
conditions, system performance of the composite link can have high
error floors due to the false estimates of FSO link. The system
performance can be improved by employing error control coding
techniques. One simple solution employing a convolutional encoder
and Viterbi decoder pair is also analyzed in this thesis. Another
important system parameter that is the average channel capacity of
the FSO system is analyzed under weak and moderate turbulence
conditions. Theoretical derivations that are verified via
simulation results indicate a reliable high data rate
communication system that is effective in long distances.

\end{abstract}
%
%
\begin{ozet}
Bu çalışmada karma darbe-radyo ultra-genişbant haberleşme sistemi
önerilmiştir. Önerilen sistem modeli ultra-genişbant sinyalleri
optik serbest uzay (OSU) üzerinden uzun mesafelerde taşıması
amaçlanmıştır ve sistem tasarımına bağlı olarak elektriksel OSU
kestirimleri direkt olarak kullanılabilir veya son kullanıcıya
radyo frekansı (RF) üzerinden kısa mesafelerde dağıtım yapabilir.
Ancak, OSU gönderim kanallarındaki türdeşsizlikler alınan sinyalin
yoğunluğunda dalgalanmalara neden olur ve atmosferdeki türbülans
kaynaklı bu etkiler sistem performansını yakından etkilemektedir.
OSU kanalları tanımlamak için deneysel ölçümlerden çıkarılan
belirli istatistiksel dağılımlar mevcuttur ve bu modelleri
kullanarak önerilen sistemin hata olasılık analizi zayıf, orta ve
sert türbülans bölgelerinde incelenmiştir. Yapılan teorik analiz
sonuçlarına göre atmosfer koşullarına bağlı olarak sistem
performansında yanlış OSU kestirimlerinin getirdiği yüksek hata
tabanları gözlemlenmiştir. Sistem performansının iyileştirilmesi
için hata kontrol kodlaması önerilmiştir. Örnek olarak, OSU
sisteminde gönderici tarafında evrişimli kodyalıcı ve alıcı
tarafında Viterbi kodçözücüsü kullanıldığı durum incelenmiş ve bu
durumun hata olasılık analizi yapılmıştır. Önemli bir başka sistem
parametresi, ortalama kanal kapasitesi OSU kanallar için zayıf ve
orta türbülans koşullarında incelemiştir. Simulasyon sonuçları ile
doğrulanan teorik çıkarımlar, önerilen sistemin uzun mesafelerde
güvenilir bir şekilde yüksek veri hızlarına çıkabileceğini
göstermiştir.

\end{ozet}

\tableofcontents \listoffigures \listoftables

\begin{symabbreviations}
\sym{$A_0$}{Path loss at a reference distance of 1 meters}
\sym{$C_n^2$}{Wavenumber spectrum structure parameter} \sym{D}{The
link distance} \sym{$D_0$}{Diameter of the receiver aperture}
\sym{$d_0$}{Correlation length of the atmospheric turbulence}
\sym{$E \left[ \cdot \right]$}{Expectation operation}
\sym{$E_{RX}$}{Energy of the received pulse} \sym{$E_{TX}$}{Energy
of the transmitted pulse} \sym{$F\{\cdot\}$}{Fourier transformation
operation}\sym{$f_C$}{Center frequency of the
transmitted message}\sym{$f_H$}{Higher frequency of the -10 dB
emission point}\sym{$f_L$}{Lower frequency of the -10 dB emission
point} \sym{$G$}{Average total multi-path gain}
\sym{$G^{m,n}_{p,q}$}{Meijer's G-function} \sym{$G_0$}{The
reference power gain evaluated at 1 meters} \sym{$g_0$}{Mean of
the random variable $g$}\sym{$h_F$}{FSO channel}\sym{$h_R$}{RF
channel} \sym{$I$}{Light intensity} \sym{$I(X;Y)$}{Mutual
information between random variables denoted by $X$ and $Y$}
\sym{$\overline{I}$}{Mean of light
intensity}\sym{$K_n(\cdot)$}{$n^{th}$ order modified Bessel
function of second kind} \sym{$k$}{Wavenumber} \sym{L}{Link
distance} \sym{$L_0$}{Outer scale of turbulence} \sym{$l_0$}{Inner
scale of turbulence} \sym{m}{Meters} \sym{$nm$}{Nanometers}
\sym{$n(\overrightarrow{r},t)$}{Refractive index}
\sym{$n_0$}{Average refractive index}
\sym{$n_1(\overrightarrow{r},t)$}{Fluctuation component induced by
spatial variations of both pressure and temperature}
\sym{$PL_0$}{Path loss at a reference distance of 1 meters}
\sym{$p_{jk}$}{Bernoulli random variable} \sym{$T_j$}{Arrival time
of the $j^{th}$ cluster} \sym{$v$}{Root mean square of the wind
speed}\sym{$X$}{Random variable denoting log-amplitude
fluctuations of the atmospheric turbulence
channel}\sym{$x_F$}{FSO-UWB TH-PPM pulse} \sym{$x_R$}{RF-UWB
TH-PPM pulse} \sym{$z_0$}{Effective height of the turbulent
atmosphere} \sym{}{}
\sym{$\alpha$}{Random variable denoting small scale scatters}
\sym{$\alpha_{jk}$}{The channel coefficient of the $j^{th}$
cluster $k^{th}$ multipath} \sym{$\beta$}{Random variable denoting
large scale scatters} \sym{$\beta_{jk}$}{The log-normal
distributed channel coefficient of $j^{th}$ cluster $k^{th}$ path}
\sym{$\delta \left( \cdot \right)$}{Dirac-delta function}
\sym{$\eta$}{Electro-optical conversion coefficient}
\sym{$\Gamma$}{Power decay factor for clusters}
\sym{$\Gamma(\cdot)$}{Gamma function}
\sym{$\Gamma_n(\cdot)$}{Autocorrelation function of $n$}
\sym{$\kappa$}{Log-normal fading random variable}
\sym{$\kappa_{jk}$}{Gaussian random variable}
\sym{$\Lambda$}{Poisson arrival rate of clusters}
\sym{$\lambda_1$}{Poisson arrival rate of multipaths within
clusters} \sym{$\lambda$}{Wavelength} \sym{$\mu_{jk}$}{Mean of the
random variable $\kappa_{jk}$} \sym{$\Omega_0$}{Normalization
factor of the total received
power}\sym{$\Phi_n(\cdot)$}{Wavenumber spectrum}
\sym{$\sigma_g^2$}{Variance of the channel amplitude gain}
\sym{$\sigma_{I}^2$}{Variance of log-amplitude fluctuations in
terms of scintillation index} \sym{$\sigma_{SI}^2$}{Scintillation
index}\sym{$\sigma_X^2$}{Variance of log-amplitude fluctuations}
\sym{$\sigma_{\xi}$}{Variance of the fluctuations of the channel
coefficient for clusters} \sym{$\sigma_{\zeta}^2$}{Variance of the
fluctuations of the channel coefficients for rays within clusters}
\sym{$\sigma_1^2$}{Rytov variance for plane waves}
\sym{$\sigma_2^2$}{Rytov variance for spherical waves}
\sym{$\xi_{jk}$}{Channel coefficient of fluctuations on each
cluster} \sym{$\tau_{jk}$}{The arrival time of the $j^{th}$ paths
(rays) of the $k^{th}$ cluster} \sym{$\zeta_{jk}$}{Channel
coefficient of fluctuations on each contribution}\sym{}{}

\sym{AWGN}{Additive white Gaussian noise} \sym{BER}{Bit error
rate} \sym{BPSK}{Binary phase shift keying}\sym{CD}{Compact disc}
\sym{CEPT}{European Conference of Postal and Telecommunications}
\sym{CLT}{Central limit theorem} \sym{CM1}{Channel model
1}\sym{DVD}{Digital versatile disc}\sym{DEP}{Detection error
probability} \sym{ECC}{Electronic Communications Committee}
\sym{EDFA}{Erbium-doped fiber amplifier}\sym{EIRP}{The equivalent
isotropic radiated power} \sym{FCC}{Federal Communications
Commission in the United States} \sym{FPLD}{Gain-switched
Fabry-Perot laser diode (FPLD)} \sym{FSO}{Free-Space Optics}
\sym{GPS}{Global Positioning System} \sym{IEEE}{Institute of
Electrical and Electronics Engineers} \sym{IR}{Impulse Radio}
\sym{ITU}{International Telecommunications Union}
\sym{LOS}{Line-of-sight}\sym{ML}{Maximum-likelihood}\sym{MLSD}{Maximum-likelihood
sequence detection} \sym{NLOS}{Non-line-of-sight}\sym{OOK}{On-Off
keying}\sym{PAN}{Personal Area Network} \sym{PDF}{Probability
density function}\sym{PPM}{Pulse position
modulation}\sym{RF}{Radio frequency}\sym{RMS}{Root mean
square}\sym{SNR}{Signal-to-noise ratio}\sym{S-V}{Saleh
Valenzuela}\sym{TF}{Tunable
Filter}\sym{TH}{Time-hopping}\sym{UWB}{Ultra-Wideband}\sym{WDM}{Wavelength
division multiplexing}\sym{WLAN}{Wireless Local Area
Networks}\sym{WPAN}{Wireless Personal Area Networks}
\end{symabbreviations}

\chapter{INTRODUCTION}
\pagenumbering{arabic} Ultra-wideband (UWB) systems characterize
those that employ very narrow pulses in time usually on the order of
nanoseconds to convey information. Consequently, these pulses
occupy very large bandwidths in frequency domain. Due to their
characteristics, UWB systems have unique attractive features
introducing new advances in several wireless communications fields
such as radar, imaging, networking and positioning systems. These
aforementioned fields benefit from several advantages offered by
UWB systems such as enhanced ability to penetrate through
obstacles such as concrete walls, ultra high precision ranging,
possibility to transmit at very high data rates and support
multi-user, relatively small size and less power consumption
\cite{giannakis}. Until 2001, the use of UWB systems were
limited solely to military applications and with the regulations
introduced by the Federal Communications Commission (FCC) in the
United States, these new features of UWB systems have gained its
attraction in several areas \cite{fcc}. The FCC in the United States has
allocated a huge unlicensed frequency spectrum ranging from 3.1
GHz to 10.6 GHz for UWB systems. These pulses need occupy at least
500 MHz of bandwidth within the allowed spectrum or a fractional
bandwidth of more than 20 percent where the fractional bandwidth
defines the ratio of bandwidth $B$ over the center frequency
$f_C$. Since UWB systems do not employ any carrier to transmit the
signals, the definition of center frequency needs to be clarified.
The lower and higher frequencies of $-10$ dB emission
points on the power spectrum of the transmitted signal are denoted
by $f_L$ and $f_H$ and their arithmetic mean determines the center frequency of the signal,
$f_C = (f_L + f_H)/2$. Also, the bandwidth of the signal is defined as the
difference of these two frequencies, $B = f_H-f_L$ \cite{giannakis}.

Although wireless propagation channels have been studied for more
than 50 years and several mathematical models are introduced in
literature to describe their propagation characteristics, the FCC
in the United States have accepted the IEEE 802.15.3a channel
model for the indoor and outdoor UWB channels with the
contribution from Intel Corporation in 2002. This regulation for
UWB systems are primarily developed for Wireless Personal Area
Networks (WPAN) and to enable the coexistence of UWB systems with
the current technologies, namely, the Bluetooth and the IEEE
802.11 wireless local are networks (WLAN), the FCC in the United
States has defined the spectral mask that every UWB device needs
to accompany which is presented in the next chapter.

Although UWB systems bring their unique advantages, they are
limited to short distances, usually up to 10 meters, due to severe
attenuation of these narrow pulses. To overcome this short range
limitation in radio frequency (RF) links, free-space optical (FSO)
systems have emerged as a cost efficient solution. Compared to RF
systems, FSO links offer much larger bandwidth and capacity, less
power consumption, more robust against eavesdropping and better
protection against interference \cite{qary}. Although FSO systems
are mainly used in inter-satellite and deep space communications,
these systems have attracted considerable attention to solve the
last-mile solution in the past decade \cite{kia}.

Contrary to other emerging alternatives to RF links, theoretical
background of FSO systems are known and they are very similar to
the fiber-optical communication systems. In these systems,
information can be conveyed in the intensity, frequency, phase or
polarization of the signal. Among these attributes, intensity is
typically used to carry information due to its simplicity and this
type of modulation employing variations in intensity is called
intensity modulation. Because of the complexities associated with
phase or frequency of the received signal, these type of
modulation are not preferred in most FSO systems \cite{zhu}.

Along with the power regulations concerning eye safety, FSO system
performance are also limited by the atmospherical effects. The
inhomogeneities in the atmosphere cause fluctuations in the
amplitude and phase of the received signal over long distances and
these effects severely degrade the performance of the system.
Besides the inhomogeneities in the atmosphere, aerosol scatterers
such as rain, snow and fog and building-sway as a result of wind
loads, thermal expansion and weak earthquakes are the other
factors degrading the system performance \cite{zhu,uysal}. For
instance, under severe fog conditions, it has been reported that
the signals attenuate on the order of hundreds of decibels per
kilometer \cite{letzepis}.

\section{Research Overview and Contributions}
This work addresses the short range coverage problem of UWB
systems that operate in RF links. To overcome this bottleneck, an
alternative technique is proposed and UWB signals are transmitted
through FSO links over several kilometers in the last mile. In FSO
systems, the impairments caused by the atmospheric turbulence
closely effect the system performance on the transmission path and
therefore, there has been extensive studies and models proposed to
describe these conditions and these can be found in
\cite{karp_commun}-\cite{majumdar}. Recent studies using these
models investigate the performance of FSO links under atmospheric
turbulence effects such as those in
\cite{kia,zhu,uysal,jain,coded_fso}. In their pioneering work, Zhu
and Kahn have applied maximum-likelihood sequence detection (MLSD)
technique to mitigate the effects of atmospheric turbulence at the
cost of increased complexity \cite{zhu}. They have also included
maximum-likelihood detection under the case when spatial diversity
is available. In \cite{uysal}, the authors have analyzed the
performance of FSO system for both correlated and independent
channels under weak turbulence conditions. The performance
analysis of a FSO system that use pulse position modulation scheme
under weak turbulence conditions is presented in \cite{jain} and
in their system model, the authors employ avalanche photodiode to
receive the signals. In his study, Kiasaleh carries out the same
analysis for the same type of photodetector but this time, the
analysis includes the effects of moderate turbulence conditions
\cite{kia}. Another recently published work in \cite{coded_fso},
the authors investigate the system performance under a more
accurate statistical distribution to model the moderate weather
conditions. Besides error analysis, the ergodic channel capacity
of FSO links in the weak turbulence regime is derived in
\cite{uysal_capacity} and for the moderate regime, its closed-form
expression is presented in \cite{gamma_nista}. Also, Hranilovic
et. al derive outage capacity analysis of an FSO system
considering the effects of pointing errors in
\cite{outage_hranilovic} and Uysal et. al apply a relay system to
their FSO system model in \cite{relay}.

However, this work focuses solely on conveying UWB signals through
FSO links over long distances and to this end, several system
models are proposed. In the first system model, the generated
optical UWB signals can be transmitted over FSO links over long
distances and at the receiver the detected electrical outputs can
be used right away or distributed to the end-user within the
building using ethernet or fiber cables at very high-speeds. It
should be emphasized again that FSO systems can achieve very high
data speeds that are compatible with fiber optical and ethernet
cables can support. As an extension of the first case, the second
scenario considers the case that the UWB signals are again
conveyed over FSO links in the last-mile but this time the
electrical outputs are distributed to the end-users by the RF link
in the last-step by employing a simple UWB transmitter and
receiver pair. Third scenario uses error-control techniques in the
FSO link to reduce the possible errors induced by the turbulence
effects. Although under good weather conditions the system
performance can achieve superior results, as the link visibility
decreases or weather conditions gets worse, error-control coding
is a necessity to endure the availability of the link and improve
the quality of service of the system. For all three scenarios,
corresponding system models are presented and their analytical
results for the detection error probability under different
turbulence conditions are analyzed. These theoretical results are
verified via simulations that confirm the presented results. To
sum up, the biggest contribution of this thesis is the complete
error analysis of the aforementioned system designs that enable
long distance transmission of UWB signals through FSO links. This
contribution is valuable because UWB systems suffer from their
short range limitations and the proposed models in this thesis
would help to overcome this bottleneck.

\section{Organization of the Thesis}
Chapter \ref{section2} provides brief information about
impulse-radio ultra-wideband systems. It presents several factors
that are considered to model channel distributions in RF links and
introduces the spectral mask proposed by several regulatory
institutions in Europe and United States that enables the
coexistence of these devices with the other equipments in the same
frequency spectrum. In its first section, several sources of
fading are discussed in detail and particular examples are given
from daily life. In the next section, the accurate distribution to
model the RF-UWB links are discussed. This accepted model is
mainly based on Saleh-Valenzuela model with some modifications.
Parameters definitions and suggested values of these parameters
are also presented towards the end of the chapter. Chapter
\ref{section3} focuses on wireless optical systems. It discusses
several commonly used modulation techniques in these systems and
introduces some conditions on the reception of the transmitted
signal along with some important parameters that determine the
link quality. Its first section explains the characteristics of
FSO channel in detail and the last section of the chapter
introduces the statistical models for different turbulence
conditions. Chapter \ref{section4} discusses the proposed system
model in this thesis. Along with the devices used in the system
model, their mathematical interpretations are also given.
Detection error probability analysis of the several possible
scenarios carried out. For instance, closed form expression are
derived for the scenario where the FSO link is used to convey the
information over several kilometers and the RF link to deliver the
signal to the end-user in the last-step and simulations are made
to verify theoretical results. This chapter is concluded with the
channel capacity analysis of the FSO links. Finally, the last
chapter addresses several concluding remarks and suggestions for
possible future work.

\chapter{IMPULSE-RADIO ULTRA-WIDEBAND SYSTEMS}\label{section2}
In wired channels such as coaxial cables, wave propagation is
predictable and usually stationary whereas in radio channels, the
behavior of wave propagation extremely random in nature
\cite{rappaport}. There are several sources that effect
electromagnetic wave propagation in radio frequency (RF) channels
such as absorption, diffraction, reflection and scattering which
all gives rise to multiple paths. On the path between transmitter
and receiver, the objects are highly likely to be absorb,
diffract, reflect or scatter the energy of the waves causing the
signals to travel by various amplitudes, phases and paths before
it reaches to the receiver. The propagation of these waves (echoes
of the original transmitted signal) are called multipath
propagation and the phenomenon of amplitude and phase fluctuations
in the received signal due to multipath propagation is called
multipath fading \cite{kohno}. Since each wave travels different
lengths of paths, pulses arrive at the receiver at different times
with different amplitudes and phases. Therefore, to describe the
behavior of the channel in different environments, several channel
models are proposed. These proposed models are mostly based
results measured from experimental data.

Several multipath components arrive to the receiver and
collectively determine the amplitude and phase of the received
signal. The first statistical model which is still widely used is
Rayleigh fading channel model and it applies to the conditions
when the bandwidth of the signal is small and the delays of the
individual multipath components do not interfere with each other
in the received signal model. This signal model assumes that there
are almost no significant scatterers nor dominant signal
reflectors in the channel. The other widely accepted channel model
is Ricean fading channel and it considers the case when the
channel exhibits fixed scatters or signal reflectors in the
channel \cite{proakis}.

Although these models are sufficient to characterize channels for
narrow-band systems, more accurate description to capture the
behavior of the multipath components is needed to characterize
systems that occupy ultra-wide bandwidths. For this purpose,
several studies have been made. In 2002, the IEEE has established
a standardization group, IEEE 802.15.3a to a develop standard for
UWB Personal Area Networks (PANs). They were to select the
multiple access schemes, modulations and a commonly agreed channel
model along with the other necessity of the standard. To this end,
IEEE 802.15.3a Task Group had been formed and established a
standard channel model to be used for the evaluation of PAN
physical layer proposals. The Federal Communications Commission
(FCC) in the United States has also decided on the data rates for
the standard model under the regulated power levels below $-41.3$
dBm in Part 15 rules to coexist with the already available
services such as Global Positioning System (GPS) and the IEEE
802.11 wireless local are networks (WLANs). IEEE 802.15.3a
standard model aims to transmit at data rates up to 100 Mb$/$s at
10 meters, 200 Mb$/$s at 4 meters and higher data rates at smaller
distances \cite{foerster2}.

The allocated spectrum to transmit these signals ranges from 3.1
GHz to 10.6 GHz and this ultra-wide bandwidth allows for several
bandwidth-demanding low power applications in wireless
communications such as short-range high-speed Internet access,
vehicular radar and sensors, asset and personnel tracking, imaging
through steel and walls, surveillance and medical monitoring
applications \cite{giannakis}. It should be emphasized that UWB
signals are on the order of nanoseconds scale so that their power
spectral density occupies such bandwidths without the need for any
carrier. Depending on the pulse width, these signals can be
designed to occupy all the 7.5 GHz spectrum in a single-band or
several multi-band UWB signals can be created such that each
occupies at minimum bandwidth of 500 MHz. The former is called as
impulse-radio ultra-wideband (IR-UWB) signalling and the latter is
referred to as multi-band UWB signalling where each has its own
consequences and this work focuses solely on IR-UWB signalling.

\begin{figure}[hb!]
\begin{centering}
\includegraphics[scale = 0.45]{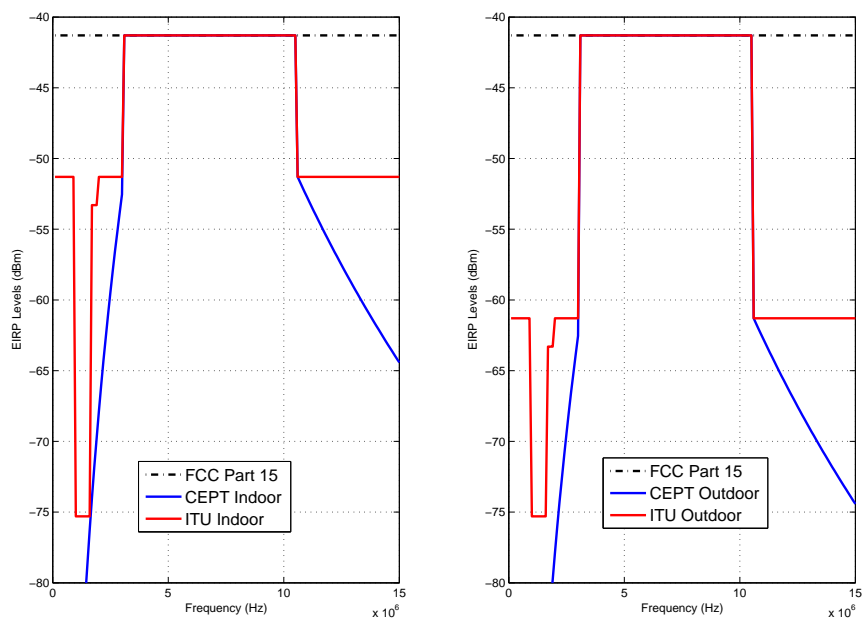}
\caption{Several indoor and outdoor spectral masks in which UWB
equipments need to satisfy.} \label{spectral mask}
\end{centering}
\end{figure}

On the other hand, UWB technology faces several challenges such as
bit error rate, capacity and throughput limitations due to
low-power regulations, network flexibility, immunity to
synchronization and many more. Along with these challenges, the
FCC has released a spectral mask that limits the equivalent
isotropic radiated power (EIRP) spectrum density of the UWB pulses
to fully coexist with the other aforementioned technologies and
this is shown in Fig. \ref{spectral mask} and Tables \ref{FCCmask}
and \ref{ITUmask} along with the regulations introduced by
European Conference of Postal and Telecommunications (CEPT) and
International Telecommunications Union (ITU). Pulse shaping
filters are commonly used to generate such pulses and several
optimal pulse shapers satisfying the UWB spectral mask are
presented in \cite{giannakis2}.

\begin{table}[h!]
\addtolength\belowcaptionskip{8pt} \caption{CEPT regulations for
indoor and outdoor UWB equipments}
     \begin{tabular}{c | c c c}
        \hline  & $f<3.1$  (GHz)& $3.1 < f <10.6$ (GHz)&
        $f> 10.6$ (GHz)
        \\ [0.5ex] \hline
        Indoor (dBm) & $-51.3 + 87 \log\left(f/3.1\right)$ & $-41.3$ & $-51.3 + 87 \log\left(10.6/f\right)$ \\
        Outdoor (dBm)
        & $-61.3 + 87 \log\left(f/3.1\right)$ & $-41.3$ & $-61.3 + 87 \log\left(10.6/f\right)$ \\
        \hline
    \end{tabular}
    \label{FCCmask}
\end{table}
\vspace{.1in}

\begin{table}[h!]
\addtolength\belowcaptionskip{8pt} \caption{ITU regulations for
indoor and outdoor UWB equipments} \centering
     \begin{tabular}{c| c| c}
        \hline Frequency (GHz) & Indoor (dBm)& Outdoor (dBm) \\\hline
        $0.96-1.61$  & -75.3 & -75.3 \\
        $1.61-1.99$ & -53.3 & -63.3 \\
        $1.99-3.1$ & -51.3 & -61.3 \\
        $3.1-10.6$ & -41.3 & -41.3 \\
        Above $10.6$ & -51.3 & -61.3 \\
        \hline
    \end{tabular}
    \label{ITUmask}
\end{table}

\vspace{.1in}

In the following sections, several sources of multipath fading in
radio frequency (RF) channels and IEEE 802.15.3a standard channel
model is be discussed in detail.

\section{Sources of Multipath Fading}
\subsection{Reflection and Propagation}

\begin{figure}[ht!]
\begin{centering}
\includegraphics[height = 1.8in]{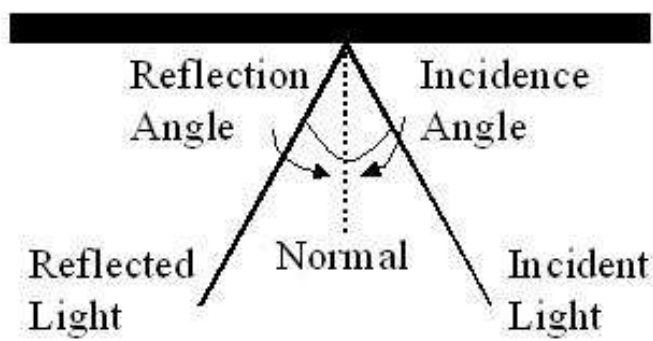}
\caption{Reflection of a wave incident on a medium.}
\label{reflection}
\end{centering}
\end{figure}

This phenomena happens at an interface between two different
(dislike) media. When waves pass from one medium to another, the
wave front suddenly changes its direction from which it
originated. This is called reflection. Reflections may be specular
(mirror-like) or diffuse (i.e. retaining only the energy, not the
image) depending on the nature of the interface. For instance,
when waves pass from a dielectric medium to a conductor medium
(dielectric-conductor), the phase of the reflected wave is
retained whereas depending on the incident angle, it may or may
not be retained when waves pass from a dielectric medium to a
dielectric medium (dielectric-dielectric).

\subsection{Diffraction} The other phenomena that propagating
waves experience is diffraction. It is the spreading out of waves.
All waves tend to spread out at the edges when they pass through a
narrow gap (for distances that are comparable to the wavelength of
the electromagnetic wave) or encounters an object. Instead of
saying that the wave spreads out or bends round a corner, it is
commonly referred as the wave is diffracted around the corner.

Intuitively, the longer the wavelength of a wave, the more it will
diffract. Another important remark is that the diffraction loss
increases with increasing frequency. A familiar example from daily
life is the reflection of image on a compact disc (CD) or digital
versatile disc (DVD). Here, the surface of a CD/DVD acts as a
diffraction grating for the reflection of images incident on its
surface. This is the main reason that when one looks at the back
of a CD/DVD, the tracks of it act as a diffraction grating and
produce iridescent reflections of the sunlight. Another example
from daily life is the color of spider webs.

\begin{figure}[ht!]
\begin{centering}
\includegraphics[height = 1.9in]{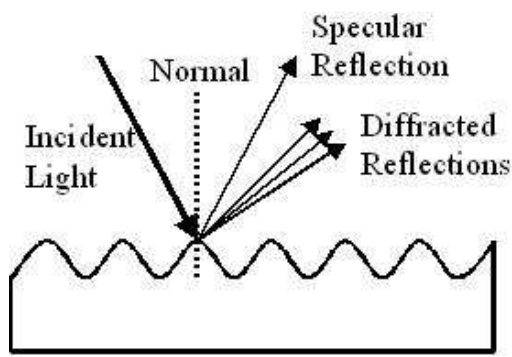}
\caption{Diffraction of a wave from a medium with curvature ends.}
\label{diffraction}
\end{centering}
\end{figure}

\subsection{Scattering}
The obstacles or inhomogeneities in the medium force to deviate
the propagation of light from its straight trajectory. When the
light is scattered by one large scatterer, then it is called
single scattering and similarly, when there are several sources of
scattering in the medium that interact with the light, then it is
named as multiple scattering. Mathematically, single scattering is
treated as random since the location of the single scatterer in
the propagation path is not known whereas multiple scattering is
considered to be deterministic since the random nature of several
scatterers can be averaged out. An example for single scattering
is when an electron is sent to an atomic nucleus. The exact
position of the nucleus relative to the electron is not known and
this bring the random nature of this interaction. To solve these
type of problems, in most cases, probability distributions are
used to describe single scattering. A light passing through a
thick fog can be given as an example for multiple scattering.
Several models are proposed in literature to model scattering
coefficients depending on the environment \cite{sources_of_fading}
such as Rayleigh scattering where the inhomogeneities in the
medium are small compared to the wavelength of the light, Mie
scattering where both sizes are comparable to each other and
non-selective scattering where the obstacle size is much larger
than the wavelength of the light. Rayleigh scattering is the
reason that the sky is blue in the day time and reddish in the
sunset. Mie scattering occurs on the lower portions of the
atmosphere where the clouds overcast the sky and the abundant
particles are large. Non-selective scattering is the reason that
the fog and clouds look white since each color of the light is
scattered around in equal portions resulting in white color.

\begin{figure}[h!]
\begin{centering}
\includegraphics[height = 2in]{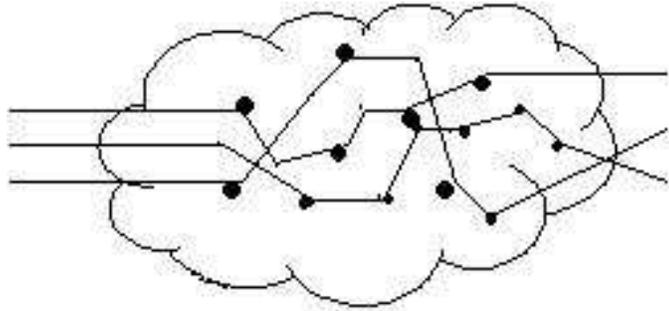}
\caption{Random scatterers in a medium effecting the propagation
path of light.} \label{scatter}
\end{centering}
\end{figure}

\section{The IEEE 802.15.3a Standard Channel Model}
This model aims to capture the multipath characteristics of
typical indoor environments where IEEE 802.15.3a devices operate.
In addition to the UWB channel measurements performed mainly in
2002, a number of measurement campaigns were carried out by the
participants of the task group \cite{foerster1}. Throughout these
campaigns, three channel models were considered in general:
Tap-delay line Rayleigh fading model, the Saleh-Valenzuela (S-V)
model and the $\Delta$-K model and these are summarized in
\cite{foerster2} and \cite{understanding}.

The experimental measurements indicate that the transmitted
signals through multipath RF environments arrive to the receiver
as clusters. The proposed UWB channel model is based on the
cluster approach first proposed by Turin and others in 1972
\cite{Turin} and further formalized by Saleh and Valenzuela (S-V
model) which was proposed in 1987 \cite{SV} only with one
significant modification. It was experimentally observed that when
lognormal distribution is used to model multipath gain magnitude
instead of Rayleigh distribution, it would better fit to
measurement data.

The S-V model which assumes the multi-path components of the same
pulse arrive as clusters, the time of arrival of these clusters
are modelled as Poisson arrival process with rate $\Lambda$ \bea
p(T_j|T_{j-1}) = \Lambda e^{-\Lambda(T_j-T_{j-1})} \label{cluster
arrival}\eea where $T_j$ and $T_{j-1}$ are the arrival times of
the $j^{th}$ and $(j-1)^{th}$ clusters, respectively. The arrival
time of the first cluster is usually set to zero, $T_{1} = 0$.
Similarly, within each cluster, consecutive multi-paths arrive
with a Poisson arrival process with rate $\lambda_1$ \bea
p(\tau_{jk}|\tau_{(j-1)k}) = \lambda_1
e^{-\lambda_1(\tau_{jk}-\tau_{(j-1)k})} \label{multipath
arrival}\eea where $\tau_{jk}$ and $\tau_{(j-1)k}$ are the arrival
times of the $j^{th}$ and $(j-1)^{th}$ paths (rays) of the
$k^{th}$ cluster, respectively. As again, the arrival times of the
first rays within each cluster is set to zero, $\tau_{j1} = 0$ for
$j = 1,...,N$. Having defined the arrival rate processes, the
proposed IEEE 802.15.3a multipath model has the following
discrete-time impulse response \bea h(t) =  \kappa \sum_{j=1}^{N}
\sum_{k=1}^{K(j)} \alpha_{jk} \delta(t-T_j-\tau_{jk}) \eea where
$\kappa$ denotes the log-normal fading random variable, $N$ and
$K(j)$ denotes the number of observed clusters and multipaths
within each cluster, respectively. $T_j$ is the delay of the
$j^{th}$ cluster and $\{\alpha_{jk}\}$ denotes the channel
coefficient of the $j^{th}$ cluster $k^{th}$ multipath. The
channel coefficients can be further can be represent by \bea
\alpha_{jk} = p_{jk} \beta_{jk} \eea where $p_{jk}$ is a Bernoulli
random variable taking equiprobable values of $\{\pm 1\}$ and it
takes account for the signal inversion due to reflection. The
other random variable, $\beta_{jk}$ is the log-normal distributed
channel coefficient that belongs to the $j^{th}$ cluster of
$k^{th}$ path. This coefficient is expressed as \bea \beta_{jk} =
10^{\kappa_{jk}/20}\eea where $\kappa_{jk}$ is a Gaussian random
variable with mean $\mu_{jk}$ and variance $\sigma_{jk}^2$ and
$\kappa_{jk}$ can further represented as \bea \kappa_{jk} =
\mu_{jk}+\xi_{jk}+\zeta_{jk} \label{Kappa}\eea where $\xi_{jk}$
and $\zeta_{jk}$ are two Gaussian random variables representing
the fluctuations of the channel coefficient on each cluster and on
each contribution, respectively. The variances of these two random
variables are represented by $\sigma_{\xi}^2$ and
$\sigma_{\zeta}^2$, respectively. The first term on the right hand
side in (\ref{Kappa}), $\mu_{jk}$ is closely related to the
exponential power decay factors of amplitude of the clusters and
multipath by \bea E\left[| \beta_{jk}|^2 \right] &=& E \left[
\left|10^{(\mu_{jk}+\xi_{jk}+ \zeta_{jk})/20} \right|^2 \right] =
E\left[|\beta_{00}|^2 \right]\exp\left(-\frac{T_j}{\Gamma} - \frac{\tau_k}{\gamma}\right) \\
\mu_{jk} &=& \frac{10\left[ \ln\left(E[|\beta_{00}|^2]
\right)-\left(\frac{T_j}{\Gamma}+\frac{\tau_j}{\gamma}\right)\right]}{\ln
10}- \frac{\ln10\left(\sigma_{\xi}^2+\sigma_{\zeta}^2 \right)}{20}
\eea where $E[|\beta_{00}|^2]$ is referred to as normalization
factor of the total received power and sometimes denoted by
$\Omega_0$ for simplicity. Often in communications, in order not
to amplify or attenuate the transmitted signal, the channel
coefficients are normalized to unity for each realization in the
system model and it requires that \bea \sum_{j=1}^{N}
\sum_{k=1}^{K(j)} \beta_{jk}^2 = 1 \label{sum1}\eea Once again,
let us remark that the arrival times of each cluster and each
multipath within a cluster is modelled by two Poisson process
which were previously defined in (\ref{cluster arrival}) and
(\ref{multipath arrival}).

The amplitude gain $ \kappa$ is also assumed to be a log-normal r.
v. with the relation $ \kappa = 10^{g/20}$ where $g$ is a normal
random variable with mean $g_0$ and variance $\sigma_g^2$. The
mean $g_0$ depends on the average total multi-path gain $G$ and
expressed as \bea g_0 = \frac{10 \ln G}{\ln 10} - \frac{\sigma_g^2
\ln10}{20} \eea where $\ln(\cdot)$ stands for the natural
logarithm. $G$ is dependent on average attenuation exponent $\psi$
by \bea G = G_0 / D^{\psi}\eea where $G_0$ is the reference power
gain evaluated at $D =1$. $A_0 = 10 \log_{10} (E_{TX}/E_{RX})$ is
the path loss at a reference distance $D_0 = 1$ in dB. In
literature, this is sometimes denoted as $PL_0$ and it is related
to reference value for power gain $G_0$ by \bea G_0 = 10 ^{-A_0 /
10}.\eea In 2003, based on several experimental results,
Ghassemzadeh and Tarokh suggested values for $A_0 = 47$ dB and
$\psi = 1.7$ to be used in line-of-sight (LOS) environment and
$A_0 = 51$ dB and $\psi = 3.5$ in NLOS environment \cite{tarokh}.
Using these definitions, the parameters for IEEE 802.15.3a UWB
channel model can be summarized in Table \ref{parameter
definitions}.

\begin{table}[h!]
\caption{IEEE 802.15.3a UWB Channel Model Parameter Definitions}
\begin{center}
\begin{tabular}{|c|l|}\hline Parameter & Definition \\ \hline
$\Lambda$ & Average arrival rate of clusters \\\hline
 $\lambda_1$ &
Average arrival rate of multipaths with in clusters (rays)
\\\hline
$\Gamma$ & Power decay factor for clusters \\\hline $\psi$ & Power
decay factor for rays within clusters \\\hline $\sigma_{\xi}$ &
Standard deviation of the fluctuations of the
\\ & channel coefficient for clusters \\\hline
$\sigma_{\zeta}$ & Standard deviation of the fluctuations of the
\\ & channel coefficients for rays within clusters \\\hline
$\sigma_g$ & Standard deviation of the channel amplitude gain
\\\hline
\end{tabular}\label{parameter definitions}\end{center}\end{table}

Since different in-door environments exhibit different channel
characteristics and the channel parameters need to be updated.
These different characteristics arise from several facts.
Depending on the distance between the transmitter and receiver and
whether a LOS component exists or not several channel model
scenarios are been proposed. The first scenario, channel model 1
(CM1) is valid for distances of $0-4$ meters with LOS component.
The second scenario, CM2 is valid for the same distances but it
considers NLOS cases. For distances of $4-10$ meters with NLOS
cases, CM3 is used. The final scenario, CM4 is the one that
includes extreme NLOS case. The channel model parameter values for
these four channels are given in Table \ref{parameter values}.

It should be emphasized that since this is a discrete-time impulse
response model, the time resolution of the system plays a crucial
role. Table \ref{parameter values} correspond to a time resolution
of 167 psec which corresponds to a bandwidth of 6 GHz. Therefore,
for any other time resolution, scaling in time domain is needed
for each of the channel coefficient realizations.

\begin{table}[h!]
\begin{center}
\caption{IEEE 802.15.3a UWB Channel Model Parameter Values}
\begin{tabular}{|l|llllllll|}\hline
Scenario & $\Lambda$ $(1/$ns$)$ & $\lambda_1$ $(1/$ns$)$ &
$\Gamma$ (ns)  &
$\psi$ (ns) & $\sigma_{\varsigma}$ (dB) & $\sigma_{\zeta}$ (dB) & $\sigma_g$ (dB) &\\
\hline CM1 & 0.0233 & 2.5& 7.1 & 4.3 & 3.3941 & 3.3941 & 3 &\\
LOS ($0$-$4$ m) &&&&&&&&\\
\hline CM2 & 0.4 & 0.5& 5.5 & 6.7 & 3.3941 & 3.3941 & 3 &\\
NLOS ($0$-$4$ m) &&&&&&&&\\
\hline CM3 & 0.0667 & 2.1& 14 & 7.9 & 3.3941 & 3.3941 & 3 &\\
NLOS ($4$-$10$ m) &&&&&&&&\\
\hline CM4 & 0.0667 & 2.1& 24 & 12 & 3.3941 & 3.3941 & 3 &\\
Extreme NLOS &&&&&&&&\\\hline
\end{tabular}\label{parameter values}
\end{center}\end{table}

\chapter{FREE-SPACE OPTICAL SYSTEMS}\label{section3}
Free-space optical (FSO) systems are originally created to
supplement fiber-optical cable systems and their apparatus and
techniques originate from fiber-optical systems. Digital
information in the form of binary signals are sent through
building roofs or window-mounted infrared laser diode
transmitters. Usually, the systems use narrow optical pulses to
transmit $1$'s and transmits nothing for $0$'s and this type of
modulation scheme is called On-Off Keying (OOK). Another common
modulation technique in FSO systems is pulse position modulation
(PPM) where there are $M$ pulse slots of $T_p$ duration. Depending
on the position of the pulse, typically equal energy pulses are
transmitted. In this scheme, synchronization between transmitter
and receiver plays an important role to determine the starting
positions of the pulses. Transmission efficiency is enhanced by
packetizing data which ensures that dividing traffic into packets
that can be independently sent and received. Also, similar to many
fiber-optical communications, FSO can support wavelength division
multiplexing (WDM) which allows a single optical path to carry
different signal channels ensured that each signal has a different
wavelength.

Typically, $850$ and $1550$ nm laser diodes are the common
commercially available products in today's technology. In these
systems, the information is transmitted by modulating the signal
intensity denoted by $I(t)$ which is produced in response to an
input electrical current signal and this process is called
intensity modulation. It is important to emphasize that the
information in these systems are carried in the intensity of the
transmitted signal contrary to the RF systems where either the
amplitude, the frequency or the phase of the signal or sometimes
any combination of these are used to transmit information. The
generated optical intensity is focused by a lens and released out
as a beam of light just like flashlights. Although these pulses
are focused by a lens, the power of the beam still disperses over
long distances. At the receiver, the transmitted light is focused
back onto a photodetector which generates current proportional to
the intensity incident on its photodetector area and this process
is called direct-detection. This process inside the photodetector
is analogous to square law devices for RF systems where the
integral of the amplitude square of the signal is taken over a
time period. In many cases, the generated current in the
photodetector is weak and amplification is employed in the
receiver end before it is demodulated. By this way, optical to
electrical conversion of the signals are accomplished.

\begin{figure}[h!]
    \centering
      \includegraphics[scale=0.5]{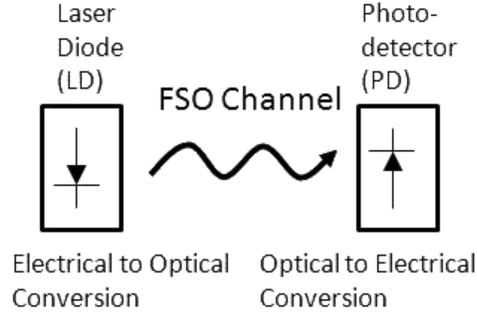}
    \caption{Electrical to optical conversion of signal via laser diode and optical to electrical conversion of the signal via photodetector through the FSO link.}\label{LDandPD}
\end{figure}

Due to the underlying structure of the channel, types of
modulation and detection of optical signals are limited to
variations on the optical intensity. This imposes the
non-negativity constraint on the signals to be transmitted and
this can be shown as \bea I(t) \geq 0, \hspace{.1in} \forall t \in
\Re\eea and the above constraint is meaningful in the sense that
the transmitted power can physically never be negative \cite{woc}.
Here, it should be emphasized that the received signal in the
photodetector that is shown by \bea y(t) = \eta I(t) x(t) + n(t)
\eea where $x(t)$ is the modulating data and $n(t)$ denotes the
noise in the system, can take negative values depending on the
noise characteristics.

Another important physical characteristics in FSO systems is beam
divergence. The transmitted beam diverges by the time it arrives
at the receiver. The degree of beam spreading depends on the lens
diameter and the received energy at the collecting lens decreases
with the square of the link distance. Without investigating the
atmospheric conditions, transmit optical power, receiver
sensitivity and size of the collecting lens, all impose
constraints on the range of the communication link for a given
data rate \cite{lastmile}.

In order to increase the range of the communication link, the
diameter of the transmitting lens needs to be increased. By
employing larger lens diameters, the power incident on the
collecting lens is increased resulting in reduced beam spreading.
Although there are physical limitations on how large the lens
diameter can be, there is also another problem associated with
using large lenses. As the diameter of the lens gets larger, the
beam spreading gets narrower causing the alignment between
transmitter and receiver to get difficult. Including the building
sway effects and thermal expansion or contraction of tall
buildings, focusing may cause problems due to narrow beams.
Therefore, many commercial products employ auto-tracking
capabilities to both ends which brings additional cost and
complexity to the system. To deploy tracking capabilities to FSO
systems, movable mechanical platforms or sometimes articulated
mirrors are used to ensure correct pointing of the transmitted
beam.

Atmospheric weather conditions such as fog, rain and snow are the
other factors that limit the range of FSO links. Among these
weather conditions, FSO systems are most susceptible to fog. For
example, in moderate dense fog, the optical signal loses 90
percent of its strength every 50 meters so for a link distance of
150 meters 99.9 percent of the transmitted optical energy will be
dissipated \cite{lastmile}. Therefore, while deploying FSO
systems, weather conditions of the deployment cites need to be
carefully investigated.

To solve the range versus reliability problem for FSO systems,
they are designed with limited link lengths to  form an optical
mesh topology. By this way, when several links fail, the spider
web like structure may still ensure the signal to be redirected
through a different path ensuring availability of the service. On
the other hand, the mesh topology brings complexity to the system.
To control the distribution of the signals along the paths,
network management type systems are needed.

\section{Free-Space Optical Channels}
In order to fully design and analyze the properties of a
communication channel, the propagation characteristics of the path
from transmitter to receiver must be taken into account.
Therefore, electromagnetic propagation of the waves needs to be
investigated. There are two types of propagation depending on the
medium such as guided and unguided transmissions. In guided
channels, wave guides are used to confine the wave propagation
from transmitter to receiver. Fiber-optical cables are the primary
examples for such communication systems. On the other hand, in an
unguided channel, the transmitter releases the generated field
freely into a medium without any attempt to control its
propagation except to control its antenna gain pattern. This
unguided channel is often called as space channel and common
examples of this channel are free space, the atmosphere or the
ocean (underwater) \cite{karp_commun}. Unfortunately, the
characteristics of unguided (space) channels primarily depend on
the properties of the medium. Free-space channels are the simplest
type of unguided channels where the medium between transmitter and
receiver is free-space.

In FSO links, inhomogeneities induced by the temperature and
atmospheric pressure cause fluctuations in the received light
intensity \cite{zhu} which lead to turbulence induced fading.
Depending on the characteristics transmission path, this type of
fading may be severe and thus, limit the performance of the
communication system by increasing the link error probability.
Also, weather conditions such as fog, rain and snow can lead to
aerosol scattering and degrade the performance of FSO links and
these effects are discussed in \cite{aerosol1} and
\cite{aerosol2}.

In FSO communication systems, there are two parameters that
describe turbulence-induced fading such as the correlation length
of intensity fluctuations, $d_0$ and the correlation time of
intensity fluctuations denoted by $\tau_0$. These two parameters
closely effect the link performance by imposing constraints on
aperture diameter and bit duration. If the diameter of the
receiver aperture denoted by $D_0$ is larger than the correlation
length, $d_0$ then turbulence-induced fading can be reduced
significantly by employing multiple point receivers in the
receiver side and this technique is called aperture averaging
\cite{andrews_random_media}. Since $D_0>d_0$ can not be always
satisfied, to mitigate the effects of fading, several approaches
have been proposed in \cite{zhu}. These approaches include
temporal-domain and spatial-domain techniques. To reduce the
effects of fading by temporal-domain techniques is
maximum-likelihood (ML) symbol-by-symbol  under the assumption
that the receiver has the knowledge of marginal fading but does
not have the temporal fading correlation nor the instantaneous
fading states. When the receiver knows the joint temporal fading
distribution but no the instantaneous fading state, the receiver
can employ MLSD. As spatial-domain
techniques, multiple receivers must be employed at the receiver to
collect the incident intensity. As in radio frequency
communication systems, receivers need to be separated as far as
possible in order to maximize the receive diversity gain.
Therefore, uncorrelated turbulence induced fading can be achieved.
In practice, separating receivers may not be made possible due to
the large area they would occupy and then correlated turbulence
fading would occur. Techniques to mitigate the above mentioned
turbulence induced fading effects in temporal and spatial domain
are explained in \cite{zhu}.

An important parameter in Kolmogorov theory is refractive index
and it is defined as \cite{zhu} \bea n(\overrightarrow{r},t) = n_0
+n_1(\overrightarrow{r},t) \eea where $n_0$ defines the average
index and $n_1$ denotes the fluctuation component induced by
spatial variations of both temperature and pressure. The
autocorrelation function of $n_1$, represents the spatial
coherence of the refractive index denoted by
$\Gamma_{n_1}(\overrightarrow{r_1},\overrightarrow{r_2})$. its
Fourier transform denoted by $\Phi_n(k)$ is called wave number
spectrum and it can be represented as \bea \Phi_n(k) =
F\left\{E\left[n_1(\overrightarrow{r_1},t_1) \cdot
n_1(\overrightarrow{r_2},t_1)\right] \right\} \eea where
$F\{\cdot\}$ denotes the Fourier transform operation.

According to Kolmogorov theory, turbulence characteristics can be
divided into three regions.

\begin{itemize}
\item Input Range: This region considers the case when the eddy
size is greater than outer scale of turbulence $L_0$, ie, (eddy
size$ > L_0$). In this region, the turbulence is anisotropic and
the wind shear and temperature gradient introduces energy to the
turbulence. Since the energy sources may vary, no general formula
is applicable to this chaotic region \cite{ishimaru}.

\item Inertial subrange: This region considers the case when the
eddy size is between the inner and outer scale of turbulence, ie,
($l_0 <\mathrm{eddy \hspace{0.05in} size}<L_0$). In this region,
the turbulence is fairly isotropic and the kinetic energy of the
eddies dominates the energy due to the viscosity dissipation.

\item Dissipation range: This region considers the case when the
eddy size is smaller than the inner scale of turbulence, ie,
($l_0>\mathrm{eddy \hspace{0.05in} size}$). In this region the
energy due to the viscosity of eddies dominates the kinetic
energy.
\end{itemize}
Depending on the ranges, the Kolmogorov wavenumber spectrum can be
found differently. Unfortunately, for the input range, the
spectrum is still unknown \cite{ishimaru}. For inertial range, it
can be given by \bea \Phi_n(K) = 0.033 C_n^2 K^{-11/3}, \eea where
$C_n^2$ is the wavenumber spectrum structure parameter and it is
an altitude-dependent parameter. For dissipation range, the
spectrum is considered to be zero, ie, $\Phi_n(K) = 0$. The
following formula is often considered to characterize all three
regimes \bea \Phi_n(K) = 0.033 C_n^2 (K^2+1/L_0^2)^{-11/6}
\exp(-K^2/K_m^2) \eea where $K_m = 5.92/l_0$. This equation is
referred to as \emph{von Karman spectrum}. Using Hufnagel and
Stanley model, the wavenumber spectrum structure parameter can be
given by \cite{ishimaru} \bea C_n^2 (z) = K_0 z^{-1/3}
\exp(-z/z_0) \eea where $K_0$ describes the strength of the
turbulence and $z_0$ is the effective height of the turbulent
atmosphere. Several works in literature define the wavenumber
spectrum structure parameter as \cite{uysal,andrews_scintillation}
\begin{align} C_n^2(z) =& 0.00594(v/27)^2(10^{-5}z)^{10}\exp(z/1000)
\\& +2.7 \times 10^{-6} \exp(-z/1500)+A \exp(-z/1000) \nonumber
\end{align} where $z$ is altitude in meters, $v$ is the root mean
square (rms) of the wind speed in meters per second and A is the
nominal value of $C_n^2(0)$ at the ground and its unit is
$m^{-2/3}$. The wave number spectrum structure parameter, $C_n^2$
can vary from $10^{-17}$ $m^{-2/3}$ for weak turbulence regime to
$10^{-13}$ $m^{-2/3}$ for strong turbulence regime. A typical
average value is assumed to be $10^{-15}$ $m^{-2/3}$ in
\cite{karp_channel}.

This work assumes that the energy of large scale eddies are
redistributed without loss to eddies of decreasing size until
finally dissipated by viscosity \cite{zhu}. Therefore, the size of
turbulence eddies may vary from millimeters to meters. Denoting
the inner and outer scale of the eddies by $l_0$ and $L_0$,
respectively, when the link distance $L$ satisfies the following
condition \bea l_0 < \sqrt{\lambda L} < L_0 \eea where $\lambda$
is the wavelength of the transmitted signal, $l_0$ and $L_0$ are
defined as above, then the correlation length of the fading
channel, $d_0$ can be approximated by \cite{andrews_random_media}
\bea d_0 \approx \sqrt{\lambda L} \eea where $\sqrt{\lambda L}$
denotes the Fresnel zone of the turbulence. Here, it needs to be
mentioned that, the above equation does not take into account of
aerosol effects such as rain and snow. These effects would further
degrade the coherence of the optical field and lead to decrease in
the correlation length of the fading channel. In literature,
``frozen air" model is used while modelling the behavior of the
eddies in the atmosphere. As the name suggests, this model assumes
that the eddies present along within the path preserve their
relative location to each other.

\section{Statistical Models of Turbulence Regimes}

In FSO channels, depending on the strength of atmospheric
turbulence the distribution of light intensity is categorized into
three regimes such as weak, moderate and strong turbulence
conditions. All these conditions are represented via distinct
distributions that accurately model the channel fadings under the
associated regime. The weak turbulence is represented by
log-normal distribution, moderate turbulence is described by
gamma-gamma distribution and at the extreme case, strong
turbulence is denoted by negative exponential distribution. These
statistical models will be analyzed in detail in the following
subsections.

\subsection{Weak Turbulence Conditions}

The emitted light from the transmitter propagates through large
number of elements of the atmosphere and at times where each
element on the propagation path causes independent, identically
distributed (i.i.d.) scattering and phase delay, the central limit
theorem (CLT) can be invoked to represent the marginal
distribution of log-amplitude fluctuations that will be
represented by $X$ and its distribution can be shown as \bea
f_X(X) = \frac{1}{\sqrt{2 \pi
\sigma_X^2}}\exp\left(-\frac{(X-E[X])^2}{2 \sigma_X^2} \right)
\eea where $E[X]$ denotes the ensemble average of log-amplitude
fluctuations and $\sigma_X^2$ denotes its variance. Commonly, it
is assumed that the light intensity represented by the random
variable, $I$ is related to the log-amplitude $X$ by \cite{zhu}
\begin{align} I = I_0 \exp(2 X - 2 \mu_x). \end{align} Then, the marginal
distribution of the light intensity  can be expressed as
\begin{align} f_I(I) = \frac{1}{2I\sqrt{2\pi
\sigma_x^2}}\exp\left(-\frac{(\ln(I/I_0))^2}{8\sigma_x^2}\right).
\label{pdfweak}
\end{align} The variance of log-amplitude fluctuations of plane and spherical
waves can be found using the following equations
\cite{zhu,karp_channel} \bea \sigma_x^2|_{\mathrm{plane}} &=& 0.56
\left(\frac{2\pi}{\lambda}\right)^{7/6} \int_0^L C_n^2(x)
(L-x)^{5/6} \mathrm{d}x \label{sigmaxcn2plane}\\
\sigma_x^2|_{\mathrm{spherical}} &=& 0.56
\left(\frac{2\pi}{\lambda}\right)^{7/6} \int_0^L C_n^2(x)
\left(\frac{x}{L}\right)^{5/6} (L-x)^{5/6} \mathrm{d}x.
\label{sigmaxcn2spherical}\eea where $C_n^2$ is the wavenumber
spectrum structure parameter as described previously, and $L$ is
the propagation path between the transmitter and receiver. Also,
below expression is commonly use to relate scintillation index and
variance of log-amplitude fluctuations such that \bea
\sigma_{SI}^2 = \exp \left( \sigma_X^2\right) -1. \eea Although
above expressions can handle varying wavenumber spectrum structure
parameter, in this work, it is assumed to be constant along the
horizontal path. On the other hand, in some references such as
\cite{kia,andrews_scintillation,gamma_nista}, the log-normal
distribution of the light intensity has a similar representation
which uses the Rytov theory, and it is given as
\begin{align}f_I(I) = \frac{1}{I\sqrt{2\pi
\sigma_I^2}}\exp\left(-\frac{\ln(I)+\frac{\sigma_I^2}{2}}{2\sigma_I^2}\right)
\end{align}
where $\sigma_I^2$ denotes the aperture-averaged scintillation
index. Depending on the wave characteristics, the scintillation
index for a plane wave and spherical wave can be expressed as
\begin{align} \sigma_I^2|_{\mathrm{plane}} =& \exp\left[\frac{0.49
\sigma_1^2}{\left(
1+0.65d^2+1.11\sigma_1^{12/5}\right)^{7/6}}+\frac{0.51 \sigma_1^2
\left(1+0.69 \sigma_1^{12/5} \right)^{-5/6}}{
\left(1+0.9d^2+0.62d^2\sigma_1^{12/5}\right)^{5/6}} \right]-1 \\
\sigma_I^2|_{\mathrm{spherical}} =& \exp\left[\frac{0.49
\sigma_2^2}{\left(
1+0.18d^2+0.56\sigma_2^{12/5}\right)^{7/6}}+\frac{0.51 \sigma_2^2
\left(1+0.69 \sigma_2^{12/5} \right)^{-5/6}}{
\left(1+0.9d^2+0.62d^2\sigma_2^{12/5}\right)^{5/6}} \right]-1
\end{align} where $\sigma_1^2 = 1.23 C_n^2 k^{7/6}L^{11/6}$ and $\sigma_2^2 = 0.4 \sigma_1^2$ are the Rytov
variance for a plane wave and spherical wave, respectively, $d =
\sqrt{kD^2/(4L)}$, D is aperture diameter, $k$ is the wave number
given by $ k = 2 \pi / \lambda$, and  the parameters $C_n^2$ and
$L$ are given as above. In cases where only a single point
detector is used, ie, aperture averaging is not employed, $d$ is
taken as zero since the propagation path, $L$, is much larger than
the diameter of a single detector, $d$. Finally, although using
the relation between light intensity and Rytov variance is widely
used as in \cite{kia,gamma_nista}, it is not considered in this
work under the weak turbulence conditions but similar expressions
invoking Rytov theory under moderate turbulence conditions are
considered in following section.

In the following two figures, the close relation between standard
deviation of log-amplitude fluctuations, $\sigma_x$ and path
distance, $L$ is shown. Using (\ref{sigmaxcn2plane}) and
(\ref{sigmaxcn2spherical}), as the link distance varies from
hundred meters to a kilometer, the standard deviation of
log-amplitude fluctuations increase from $10^{-3}$ to $10^{-1}$.
This general behavior is depicted in the figures and the
wavelength of the optical transmitted wave is taken as 1550 nm and
$C_n^2$ is assumed to be constant throughout the link distance.

\begin{figure}[h!]
\begin{centering}
\includegraphics[height = 3.75in]{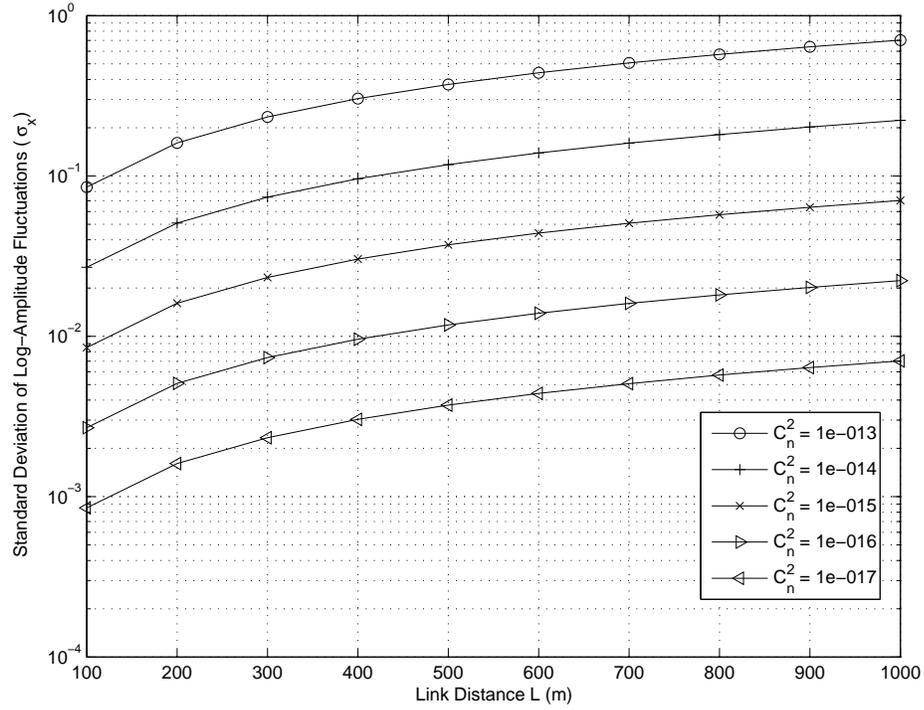}
\caption{Standard deviation of the log-amplitude fluctuations
versus propagation distance for a plane wave. The wavelength is
taken as $\lambda = 1550 \hspace{.05in} \mathrm{nm}$.}
\label{plane_wave}
\end{centering}
\end{figure}
\begin{figure}[h!]
\begin{centering}
\includegraphics[height = 3.75in]{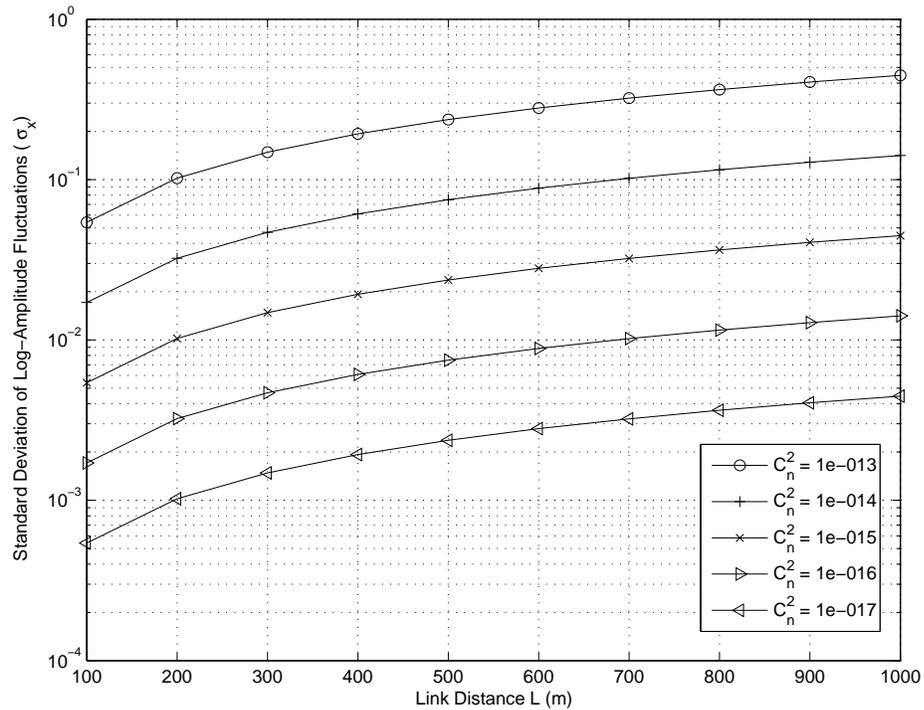}
\caption{Standard deviation of the log-amplitude fluctuations
versus propagation distance for a spherical wave. The wavelength
is taken as $\lambda = 1550 \hspace{.05in} \mathrm{nm}$.}
\label{spherical_wave}
\end{centering}
\end{figure}
\newpage

\subsection{Moderate Turbulence Conditions}
Under moderate turbulence conditions, both the small and large
scale scatterers are effective and the intensity fluctuations are
modelled by the Gamma-Gamma distribution described by the
following probability density function \bea f_I(I;\alpha,\beta) =
\frac{2(\alpha
\beta)^{\frac{\alpha+\beta}{2}}}{\Gamma\left(\alpha\right)\Gamma\left(\beta\right)}
I^{\frac{\alpha+\beta}{2}-1} K_{\alpha-\beta}\left( 2 \sqrt{\alpha
\beta I}\right) \label{gamma} \eea where $\Gamma(\cdot)$ and
$K_n(\cdot)$ are the Gamma function and the $n$th-order modified
Bessel function of the second kind, respectively \cite{bayaki}.
Small and large scale scattering effects that are represented by
$\alpha$ and $\beta$, respectively in (\ref{gamma}) are related to
the atmospheric conditions for a plane wave with aperture-averaged
scintillation index by \cite{andrews_random_media}
\bea \alpha &=& \left[\exp\left(\frac{0.49 \sigma_1^2}{\left(1+0.65d^2+1.1 \sigma_1^{12/5}\right)^{7/6}}\right)-1\right]^{-1} \nonumber\\
\beta &=& \left[\exp\left(\frac{0.51 \sigma_1^2 \left(1+0.69
\sigma_1^{12/5}\right)^{-5/6}}{\left(1+0.9d^2+0.62d^2
\sigma_1^{12/5}\right)^{5/6}}\right)-1\right]^{-1}
\label{alfabeta}\eea where $\sigma_1^2 = 1.23 C_n^2
k^{7/6}L^{11/6}$ is the Rytov variance for a plane wave, $d =
\sqrt{kD^2/(4L)}$, D is aperture diameter and the parameters
$C_n^2$, $k$ and $L$ are given as above. The same parameters are
related to atmospheric conditions for a spherical wave with
aperture averaging by
\bea \alpha &=& \left[\exp\left(\frac{0.49 \sigma_2^2}{\left(1+0.18d^2+0.56 \sigma_2^{12/5}\right)^{7/6}}\right)-1\right]^{-1} \nonumber\\
\beta &=& \left[\exp\left(\frac{0.51 \sigma_2^2 \left(1+0.69
\sigma_2^{12/5}\right)^{-5/6}}{\left(1+0.9d^2+0.62d^2
\sigma_2^{12/5}\right)^{5/6}}\right)-1\right]^{-1}
\label{alfabeta_spherical}\eea where $\sigma_2^2 = 0.4 \sigma_1^2$
(ie, $\sigma_2^2 = 0.5 C_n^2 k^{7/6}L^{11/6}$) is the Rytov
variance for a spherical wave and the other parameters, $C_n^2, d,
k$ and $L$ are as defined previously. These parameters are related
to the scintillation index by $\sigma_{SI}^2 = \alpha^{-1} +
\beta^{-1} +(\alpha \beta)^{-1}$ and moderate turbulence regime is
often assumed to be valid for $0.75 \leq \sigma_{SI}^2 < 1$.

Increase in the link distance $L$ or increase in the wavenumber
spectrum structure parameter $C_n^2$ would result in an increase
in the Rytov variance for each type of waves, $\sigma_1^2$ and
$\sigma_2^2$. Using this fact and the above equations to determine
small and large scale effects, the following figure is plotted to
depict the relation between link distance and
$(\alpha,\beta,\sigma_{SI}^2)$ parameter pairs. For small link
distances, both small and large scale scattering effect parameters
$\alpha$ and $\beta$ have large values and as the link distance
increases small scale scattering effects $\alpha$ significantly
increases after some point whereas large scale scattering effects
$\beta$ constantly decrease and approach to unity. For the
scintillation index, this parameter always increases and reaches a
saturation limit as the link distance gets larger. This saturation
regime is better described by negative exponential distribution
which is explained in the next subsection.

\begin{figure}[ht!]
\begin{centering}
\includegraphics[height = 3in]{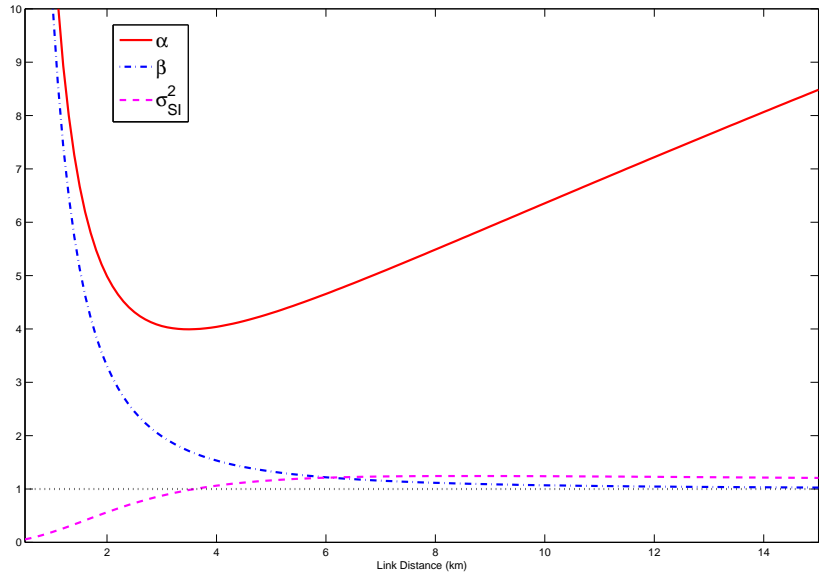}
\caption{Small and large scale scattering effects and
scintillation index as a function of link distance. The wavelength
and wavenumber spectrum structure parameter are taken as $\lambda
= 1550$ nm and $C_n^2 = 10^{-14}$ $m^{-2/3}$, respectively.}
\label{alfabetasi}
\end{centering}
\end{figure}

\subsection{Strong Turbulence Conditions}
\begin{figure}[ht!]
\begin{centering}
\includegraphics[height = 3.5in]{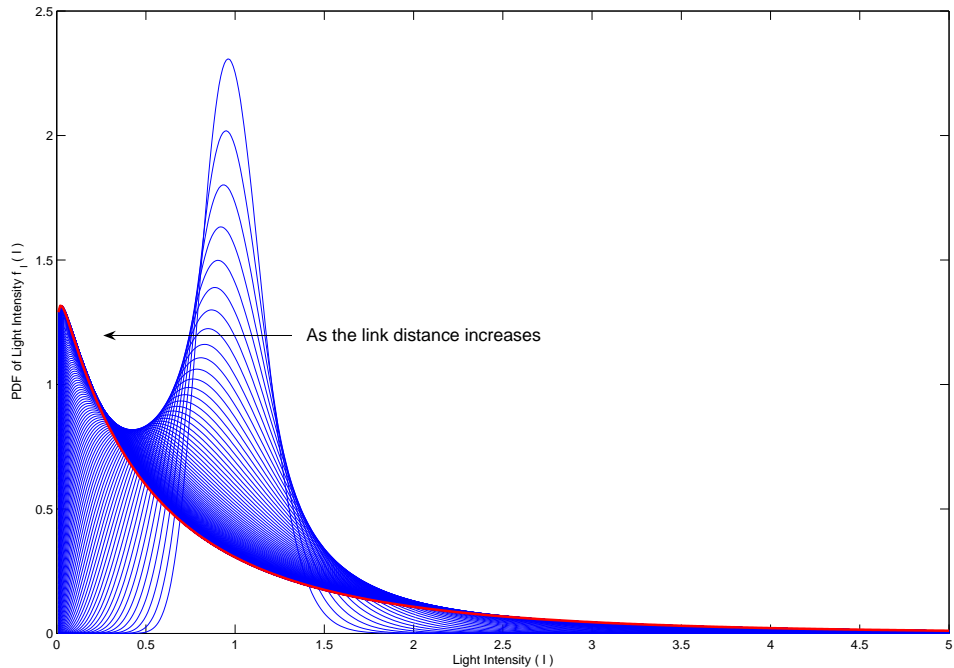}
\caption{The probability density function of several
($\alpha,\beta$) pairs as a function of link distance.}
\label{nexp fig}
\end{centering}
\end{figure}

Under strong turbulence conditions, there are many non-dominating
scatterers in the medium. Passing from medium to strong turbulence
regime corresponds to the case when the small scale scattering
effects denoted by $\alpha$ increase but the large scale
scattering effects approach represented by $\beta$ approach to
unity as the link distance $L$ increases and under this condition,
the light intensity is better modelled as one-sided negative
exponential distribution which is given by \bea f(I) =
\frac{1}{\overline{I}}\exp \left(- \frac{I}{\overline{I}}
\right)\mbox {, $I\geq0$} \label{nexp}\eea where $\overline{I}$
denotes the mean light intensity. This saturation regime
corresponds to scintillation index for the range $\sigma_{SI}^2
\geq 1$. To show that gamma-gamma distribution converges to
negative exponential distribution in the saturation regime the
following figure is plotted. For the same $(\alpha,\beta)$ pairs
in the previous subsection, the pdf's are depicted in Fig.
\ref{nexp fig}. As the link distance increases $\beta$ approaches
to unity and $\alpha$ grows unboundedly. This behavior is depicted
in the figure such that the longest link distance which is shown
as the red curve has a negative exponential distribution. Aside
from this mathematical interpretation, physically in the
saturation regime large scale effects are not dominant instead
there are several non-dominating small scale scattering effects
are effecting the propagation of light.

\chapter{PROPOSED ULTRA-WIDEBAND SYSTEM AND ITS
ANALYSIS}\label{section4}

The proposed system model in this work includes two separate
sections such as free-space optical (FSO) and radio-frequency (RF)
system models. Both of the systems employ ultra-wideband
signalling to achieve very high data rates. The FSO system is
designed such that it can take binary inputs in both electrical or
optical domain. These inputs are used to generate optical pulses
and transmitted via FSO links over several kilometers. The
received signals are demodulated in the receiver and they can be
used right away since they are converted to electrical domain
using the photodetectors in the receiver. The main advantage of
this section of the system is to reduce the time and cost to
install fiber optical cables for long distances since both the
installation time and cost to deploy FSO equipment is much shorter
and cheaper compared to deploying fiber optical equipments. Also,
depending on the link design, the received estimates of the FSO
links can be used to distribute the information inside the
buildings. Using simple and commercially available UWB transmitter
and receiver pairs, the information can be delivered to the
end-user in the last-step. Due to the severe attenuation of the
signal in the RF domain, the distance between the transmitter and
receiver in the last-step is limited to short distance up to 10
meters. For both sections of the system, their systems models are
presented and their detection error probabilities (DEP) are
analyzed in the following sections. The FSO system performance is
closely related to the atmospheric turbulence conditions.
Therefore, its performance analysis is divided into three parts
depending on the strength of the turbulence such as weak, moderate
and strong conditions. Also, simulations are carried out to verify
the derived theoretical results. Towards the end of this chapter,
channel capacity analysis of the FSO system is presented.

\section{System Models}

\begin{figure}[hb!]
\centering
\includegraphics[height=2in, width = 5in]{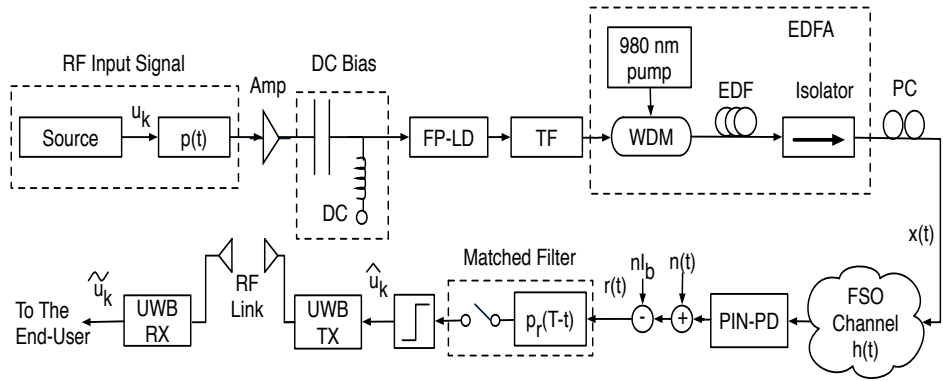} 
\caption{System model of the proposed communication system over
FSO + RF links.} \label{system_model}
\end{figure}

\subsection{Free-Space Optical System Model}

The system model shown in Fig. \ref{system_model} is considered
and the approaches presented in \cite{lin} and \cite{lin2} are
adopted. The proposed system employs gain-switched Fabry-Perot
laser diode (FPLD), a tunable-filter (TF) and an erbium-dope fiber
amplifier (EDFA) to generate wavelength-tunable optical pulses and
is primarily developed for UWB-over-fiber links. In the system
model, the binary source outputs $\{u_k\}$'s are time-hopping
pulse-position modulated (TH-PPM) via Gaussian pulse trains and
passed through a bias-tee circuit to drive the FPLD into
gain-switched operation. The $1550$-nm FPLD, which operates with a
threshold current of $18$ mA at $25^o$ C with $0.8$ nm mode
spacing is biased at $16$ mA and gain-switched at $4$ GHz. The
generated optical signal is fed to the EDFA which consists of a
$980$-nm pump laser diode that pumps $50$ mW output power to
couple an erbium-doped fiber via $980/1550$-nm wavelength division
multiplexer (WDM) and an isolator to reduce back reflections and
serves both as an external-injection source and an amplifier for
the FPLD output. The output of EDFA is passed through TF which
operates in the range from $1527$ to $1562$ nm. The central
wavelength of the TF is chosen to be close to that of FPLD output
so that the system has a single wavelength output before it is
sent to FSO channel. The generated optical TH-PPM signal can be
represented as \bea\label{x(t)} x_{F}(t) = \sqrt{E_F}
\sum_{n=1}^{\infty} \sum_{j=0}^{N_s-1} p(t-nT_d - jT_f - c_j T_c -
d_n T_p) \eea where $E_F$ denotes the pulse energy amplified by
EDFA, $p(t)$ denotes the Gaussian pulse, $\{d_n\}$ and $\{c_j\}$
are the binary information and pseudo-random code sequences for
time hopping, respectively. $T_d$, $T_f$, $T_c$ and $T_p$
represent the symbol, bit, chip and pulse durations, respectively
and bits are repeated $N_s$ times in a symbol period.

The FSO channel is described by the unit impulse response
$h(t)=I(t)+I_b$ where $I(t)$ and $I_b$ are the instantaneous light
intensity and the background radiation whose effects are removed
at the receiver as in \cite{zhu}, respectively. Also, experiments
show that the coherence time of the FSO links is large enough to
approximate the intensity as  a constant during the transmission
of a frame, i.e., $I(t)\approx I$.

The photo-detector receives the photon flux incident on the
detector area and produces a current that is proportional to the
received photons, providing the optical to electrical conversion
of the received signal. The conversion coefficient $\eta$ ($0<
\eta \leq 1$) indicates the efficiency of the photodetector. After
the removal of the background radiation bias $\eta I_b$, the
resulting received signal can be represented as
\begin{eqnarray}\label{r(t)}
\lefteqn{r_{F}(t) = \eta I x_{F}(t)  + n(t) } \\
&& \hspace{-.3in}=\eta I\sqrt{E_F} \sum_{n=1}^{\infty}
\sum_{j=0}^{N_s-1} p(t-nT_d - jT_f - c_j T_c - d_n T_p) + n_F(t)
\nonumber
\end{eqnarray} where $n_F(t)$ represents the combined effects of both
the thermal noise and the shot noise, which can jointly be
modelled as an additive white Gaussian noise  (AWGN) with zero
mean and variance $N_0/2$, i.e., $n_F(t)\sim {\mathcal N}(0,
N_0/2)$. The received signal is passed through the matched filter
$x_{MF}(t)$ which can be written as \beq\label{xr(t)} x_{MF}(t) =
\frac{1}{\sqrt{N_s}} \sum_{n=1}^{\infty} \sum_{j=0}^{N_s-1}
p(t-c_j T_c) - p(t - c_j T_c - T_p) \eeq resulting in the sampled
output \beq \label{y(nT)} y_{F}(kT_d) = \int_{(k-1)T_d}^{kT_d}
r_{F}(t) x_{MF}(t) dt = \pm \eta I \sqrt{N_s E_F}  + \nu_k. \eeq
\noindent Here, the noise term $\nu_k = \int_{(k-1)T_d}^{kT_d}
n_{F}(t) x_{MF}(t) dt$  is also AWGN with $\nu_k \sim {\mathcal
N}(0, N_0/2)$. To make decisions on the sampled output values
zero-threshold detection is employed for the matched filter
outputs. The estimates of these binary source outputs,
$\{\widehat{u}_k \}$ 's are then input to an UWB transmitter to be
transmitted to the end-user.

\subsection{Radio-Frequency System Model}
In this subsection, radio-frequency section of the system model is
considered. It should be emphasized that this part of the system
constitutes the last-step to the end-user. The TH-PPM pulse that
is generated by the UWB transmitter can be expressed as
\bea\label{xRF)} x_{R}(t) = \sqrt{E_R} \sum_{n=1}^{\infty}
\sum_{j=0}^{N_s-1} q(t-nT_d - jT_f - c_j T_c - d_n T_p) \eea where
$E_R$ denotes the energy of the pulse, $q(t)$ denotes the Gaussian
monocycle pulse train satisfying the UWB regulations. The UWB
channel is modelled as \bea h_{R}(t) = \kappa \sum_{l=1}^{N}
\sum_{k=1}^{K(l)} \alpha_{lk} \delta(t-T_l-\tau_{lk}) \eea where
$\kappa$ denotes the log-normal fading random variable, $N$ and
$K(l)$ denotes the number of observed clusters and multipaths
within each cluster, respectively. $T_l$ is the delay of the
$l^{th}$ cluster and $\alpha_{lk}$ denotes the channel coefficient
of the $l^{th}$ cluster $k^{th}$ multipath and it can be expressed
as $\alpha_{lk} = p_{lk} \beta_{lk}$ where $p_{lk}$ is a Bernoulli
random variable taking values of $\pm 1$ and $\beta_{lk}$ is the
log-normal distributed channel coefficient. To normalize each
channel realization to unity requires that \bea \sum_{l=1}^{N}
\sum_{k=1}^{K(l)} \beta_{lk}^2 = 1. \label{sum1}\eea \noindent The
amplitude gain $\kappa$ is also assumed to be a log-normal r. v.
with the relation $\kappa = 10^{g/20}$ where $g$ is Gaussian
distribution with mean $g_0$ and variance $\sigma_g^2$. The mean
$g_0$ depends on the average total multipath gain $G$ and
expressed as \beq g_0 = \frac{10 \ln G}{\ln 10} - \frac{\sigma_g^2
\ln10}{20} \eeq \noindent where $\ln(\cdot)$ is the natural
logarithm. $G$ is dependent on average attenuation exponent $\psi$
by $G = G_0 / D^{\psi}$, and $G_0$ is the reference power gain
evaluated at $D =1$. $A_0 = 10 \log_{10} (E_{TX}/E_{RX})$ denoting
the path loss at a reference distance $D_0 = 1$ in dB is related
to $G_0$ by $G_0 = 10 ^{-A_0 / 10}$. Assuming the transmission of
the Gaussian monocycle pulse train defined in (\ref{xRF)}), the
received signal can be expressed as
\begin{align} r_{R}(t) &= x_{R}(t)\ast h_{R}(t) + n(t)
\\&= \kappa \sqrt{E_R}\sum_{n=1}^{\infty}
\sum_{j=0}^{N_s-1}\sum_{l=1}^{N} \sum_{k=1}^{K(l)} p(t-nT_d - jT_f
- c_j T_c - d_n T_p-T_l-\tau_{lk}) + n_R(t).\nonumber\end{align}
\noindent Using an UWB receiver, estimates of the signal that
passed through FSO and RF links, $\{\widetilde{u}_k\}$'s are
obtained.

\section{Detection Error Probability Analysis}
In this section, the error probability of the composite system
including the introduced by both FSO and RF links is presented.
The total error probability of the complete hybrid system is
denoted by $P_e(\gamma_F,\gamma_R)$ and it can be expressed as
\begin{align} P_e(\gamma_F,\gamma_R) =
P_{e,F}(\gamma_{F})\left[1-P_{e,R}(\gamma_{R})\right]+P_{e,R}(\gamma_{R})\left[1-P_{e,F}(\gamma_{F})\right]
\end{align}
where $P_{e,F}(\gamma_{F})$, $P_{e,R}(\gamma_{R}) $ and $
\gamma_{F}$, $\gamma_{R} $ are the error probabilities and symbol
signal-to-noise ratios of the FSO and RF systems, respectively.
Since binary TH-PPM is employed, the above equation includes
errors introduced by both FSO and UWB links of the composite
system. Although the primary scope of this work is to analyze the
FSO system performance for different channel regimes but also, the
performance analysis of the RF channel is briefly investigated and
is presented along with the analysis of the FSO channel.

The detection error probability (DEP) for the binary TH-PPM signal
model over FSO subsystem assumes perfect synchronization between
the transmitter and receiver. Also, channel state information
(CSI) is assumed to be available at the receiver. Thus, the
detection error probability can be expressed as \bea
P_{e,F}(\gamma_{F},I) = \frac{1}{2} \erfc \left( \sqrt{\frac{N_s
\gamma_{F} (\eta I)^2}{2}}\right) \eea where $\erfc(\cdot)$ is the
complementary error function and $\gamma_{F} = \mathrm{E}[E_F /
N_0]$ is the average pulse signal-to-noise ratio (SNR) and
$\mathrm{E}[\cdot]$ denotes the expectation operation. Notice that
due to the random fluctuations in its amplitude, the light
intensity is modelled as a random variable whose distribution is
dependent on the turbulence region. Expectation is taken over the
distribution of the light intensity to find the average DEP such
that
\begin{align} \label{averageprobality} P_{e,F}(\gamma_{F}) =
\mathrm{E}_I \left[ P_{e,F}(\gamma_{F},I) \right] =
\int_0^{\infty} P_{e,F}(\gamma_{F},I) f_I(I) \mathrm{d}I.
\end{align}

\subsection{Error Probability Under Weak Turbulence Conditions}
Under weak turbulence conditions, the average DEP can be computed
by taking the expectation in (\ref{averageprobality}) over the
log-normal intensity distribution in (\ref{pdfweak}) such as
\begin{align} P_{e,F}(\gamma_{F}) = \int_0^{\infty}
\frac{1}{2} \erfc \left(\sqrt{\frac{N_s \gamma_{F} (\eta I)^2}{2}
}\right) \frac{1}{2I\sqrt{2\pi
\sigma_x^2}}e^{-\frac{\ln(I/I_0)}{8\sigma_x^2}}\mathrm{d}I.
\label{pbweak}
\end{align}\noindent Unfortunately closed form solution of
the above equation does not exist, but via Gauss-Hermite expansion
in \cite{handbook} which is given by \bea \int_{-\infty}^{\infty}
g(y) e^{-y^2} \mathrm{d}y  \approx \sum_{i = 1}^{n} \omega_i
g(y_i) \label{herm} \eea \noindent where $y_i$ and $\omega_i$,
$i=1,..,n$, are the $i^{th}$ root (abscissa) and associated
weight, respectively, (\ref{pbweak}) can be approximated after a
change of random variables $y=\ln(I/I_0)/\sqrt{8\sigma_x^2}$ as
\bea \label{pbweakGH} P_{e,F}(\gamma_{F}) \approx
\frac{1}{2\sqrt{\pi}} \sum_{i=1}^{n} \omega_i \erfc \left(\sqrt{
\frac{N_s \gamma_{F} \left(\eta I_0 e^{y_i \sqrt{8
\sigma_x^2}}\right)^2}{2}} \right). \eea \noindent As shown in the
simulation results, the Gauss-Hermite expansion up to $20^{th}$
order whose parameters given in \cite{handbook} provides a close
approximation to the actual integral in (\ref{pbweak}).

\subsection{Error Probability Under Moderate Turbulence Conditions}
Under moderate turbulence conditions, the average DEP can be
computed by the expectation in (\ref{averageprobality}) over the
Gamma-Gamma intensity distribution given in (\ref{gamma}) such as
\begin{align} \label{pb_moderate} P_{e,F}(\gamma_{F}) &= \int_0^{\infty}
\frac{1}{2} \erfc \left(\sqrt{\frac{N_s \gamma_{F} (\eta I)^2}{2}
}\right) \frac{2(\alpha
\beta)^{\frac{\alpha+\beta}{2}}}{\Gamma\left(\alpha\right)\Gamma\left(\beta\right)}
 I^{\frac{\alpha+\beta}{2}-1}
K_{\alpha-\beta}\left( 2 \sqrt{\alpha \beta I}\right)\mathrm{d}I.
\end{align} Using the formulation in \cite{wolf} the
integrands $\erfc(\cdot)$ and $K_n(\cdot)$ in (\ref{pb_moderate})
can be expressed in terms of Meijer's G-function and rewritten
again as \begin{align}
P_{e,F}(\gamma_{F})&=\frac{2(\alpha\beta)^{\frac{\alpha+\beta}{2}}}{\Gamma(\alpha)\Gamma(\beta)}
\int_0^\infty I^{\frac{\alpha+\beta}{2}-1}
\frac{1}{2}G^{2,0}_{0,2}\bigg[\alpha\beta I\bigg|\begin{array}{c}
-\\\frac{\alpha-\beta}{2},\frac{\beta-\alpha}{2}
\end{array}\bigg]\nonumber\\&
\times \hspace{0.1cm}
\frac{1}{2\sqrt{\pi}}G^{2,0}_{1,2}\bigg[\frac{N_s \gamma_{F} (\eta
I)^2}{2}\bigg|\begin{array}{c}1\\0,\frac{1}{2}
\end{array}\bigg] dI
\end{align}
\noindent which, using Eq. 07.34.21.0011.01 of \cite{wolf},
results in a closed-form expression such that
\begin{align}P_{e,F}(\gamma_{F}) &= \frac{2^{\alpha+\beta-3}}{\sqrt{\pi^3} \Gamma(\alpha)\Gamma(\beta)}
G^{2,4}_{5,2}\bigg[\frac{8\eta^2 N_s \gamma_{F}}{\alpha^2
\beta^2}\bigg|\begin{array}{c}\frac{1-\alpha}{2},\frac{2-\alpha}{2},\frac{1-\beta}{2},\frac{2-\beta}{2},1\\0,\frac{1}{2}\end{array}\bigg].
\end{align}
The intermediate steps in deriving the above closed-form
expression are presented in Appendix \ref{appendix error
probability}.

\subsection{Error Probability Under Strong Turbulence Conditions}
Using (\ref{averageprobality}) and the distribution given in
(\ref{nexp}), the average DEP for strong turbulence conditions can
be written as \begin{align} \label{pb_strong2} P_{e,F}(\gamma_{F})
&= \int_0^{\infty} \frac{1}{2} \erfc \left(\sqrt{\frac{N_s
\gamma_{F} (\eta I)^2}{2} }\right) \frac{1}{\overline{I}}
\exp\left( - \frac{I}{\overline{I}}\right) \mathrm{d}I \\
 &= \frac{1}{2} \left[1- \erfc\biggl(\frac{1}{\eta
\overline{I}\sqrt{2 N_s \gamma_{F}}
}\biggr)\exp\left(\frac{1}{2(\eta \overline{I})^2 N_s \gamma_{F}
}\right)\right]  \nonumber
\end{align} which again has a closed form expression.

\subsection{Error Probability in Coded FSO Links}
The system proposed in Fig. \ref{system_model} can also be
implemented with a channel encoder-decoder pair to protect the
information bits transmitted over the FSO channel against channel
effects. If a convolutional encoder is employed together with the
Viterbi decoding on the receiver side, the coded bit error
probability $P_{e,F}^{coded}\left(\gamma_F \right)$  in terms of
the uncoded error probability is expressed approximately as in pp.
531 of \cite{costello} \bea P_{e,F}^{coded}\left(\gamma_F \right)
\approx N_b \left[4 P_{e,F}\left(\gamma_F\right)
\left(1-P_{e,F}\left(\gamma_F\right)\right)\right]^{d_{free}/2}\eea
where $d_{free}$ is the minimum free distance of the convolutional
code, and  $N_b$ is the sum of the Hamming weight of all the input
sequences whose associated convolutional codeword have a Hamming
weight of $d_{free}$.

\subsection{Error Probability in RF-UWB Links}
Assuming normalized channel coefficients, i.e., (\ref{sum1})
holds, the received SNR at UWB receiver can be expressed as
$\gamma_{RF} = \kappa^2 N_s \gamma_R $, where $\gamma_R =
\mathrm{E}[E_R /N_0]$ denotes the average pulse SNR and results in
the following error probability $ P_{e,R} (\gamma_{R},\kappa)=
\frac{1}{2} \erfc\left(\sqrt{\gamma_R \kappa^2/2}\right)$.
Averaging out the pdf of $P_{e,R} (\gamma_{R},\kappa)$ over
$\kappa$ results in
\begin{align} P_{e,R} (\gamma_{R})&=\int_0^{\infty}
\frac{1}{2}\erfc\left(\sqrt{\frac{\gamma_{R} \kappa^2}{2}} \right)
\frac{20}{\kappa \ln(10) \sqrt{2\pi\sigma_g^2}}
\exp\left(-\frac{(20\log_{10} \kappa - g_0)^2}{2\sigma_g^2}\right)
\mathrm{d}\kappa. \label{rf_pb2}\end{align} This can be solved by
Gauss-Hermite expansion as in (\ref{herm}) after a change of
variables $y = (20 \log_{10}\kappa - g_0)/(\sqrt{2 \sigma_g^2})$
as \begin{align} P_{e,R} (\gamma_{R})\approx
\frac{1}{2\sqrt{\pi}}\sum_{i=1}^{N}\omega_i
\erfc\left(\sqrt{\frac{ \gamma_{R} 10^{(y_i \sqrt{2 \sigma_g^2}+
g_0)/10}}{2}} \right) \label{rf_GH}\end{align} where $y_i$ and
$\omega_i$ are as defined earlier.

\section{Simulation Results}
In this section, simulation results for the error performance of
the proposed FSO-UWB system under weak, moderate and strong
turbulence conditions and for the hybrid system including the RF
section in the last-step are presented. For the simulations, frame
size, repetition rate and the efficiency of the photo-detector are
chosen as $N = 1000$ symbols, $N_s = 2$ and $\eta =0.9$,
respectively.

\begin{figure}[h!]
\begin{centering}
\includegraphics[height=3.5in]{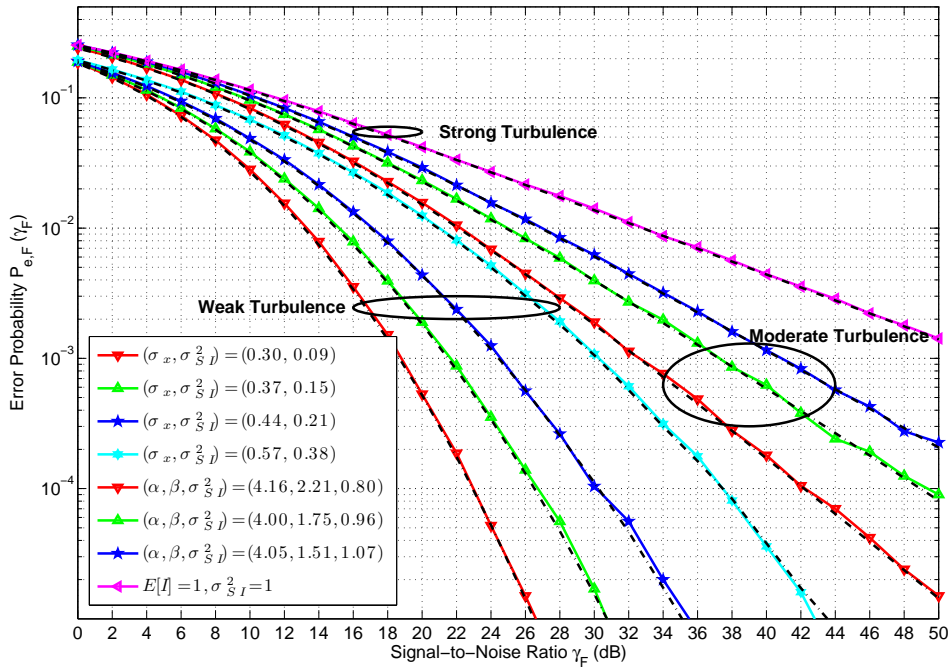}
\caption{Comparison of the simulated and theoretical DEP's for
weak, moderate and strong turbulence fading conditions.}
\label{sim_1}\end{centering}
\end{figure}

For each turbulence condition, it is assumed that $C_n^2$ is
constant through the horizontal path and for weak turbulence, it
is taken as $C_n^2 = 5.1\times10^{-15}$. For link distances over
2, 2.5, 3 and 4 km, these correspond to $(\sigma_x,\sigma_{SI}^2)
= (0.3,0.095), (0.37,0.15), (0.44,0.209)$ and $(0.57,0.380)$,
respectively. The analytical approximation derived in
(\ref{pbweakGH}) and the Monte-Carlo simulations are shown in the
group of first 4 BER curves in Fig. \ref{sim_1}. As seen in the
figure, the theoretical derivations, represented with dashed-lines
show a good correspondence to the simulated performances, depicted
in solid lines.

For moderate turbulence condition, the wavenumber spectrum
structure parameter is taken as $C_n^2 =1.76\times10^{-14}$ and
consider link distances over 2, 2.5 and 3 km. Using the relation
in (\ref{alfabeta}), the $(\alpha,\beta,\sigma_{SI}^2)$ parameter
pairs are $(4.16,2.21,0.8)$, $(4.00,1.75,0.96)$ and
$(4.05,1.51,1.07)$, respectively. The simulation results in Fig.
\ref{sim_1} for all three cases are in full accordance with the
exact closed form expressions given in terms of the Meijer's
G-function.

For strong turbulence, unit mean light intensity is assumed,
$\overline{I} = E[I] = 1$ and $\sigma_{SI}^2 = 1$. The analytical
and the Monte-Carlo simulated results are shown in the figure and
as expected, the system exhibits the worst performance in this
case due to the saturation characteristics in this regime.

\begin{figure}[h!]
\begin{centering}
\includegraphics[height=4in]{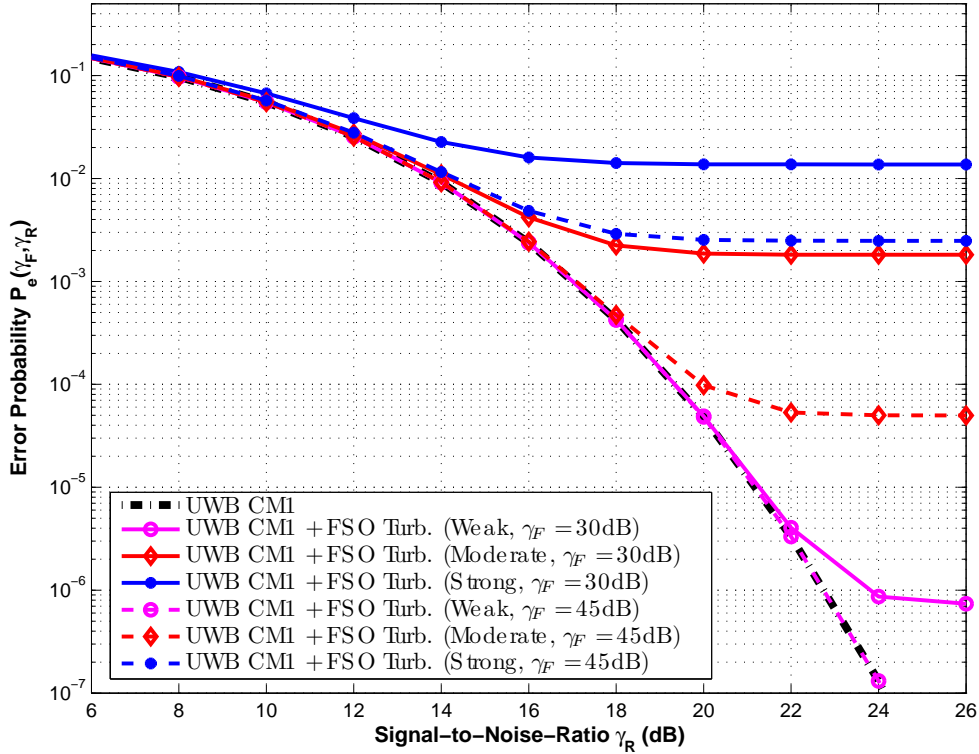}
\caption{FSO + RF UWB System Performance of a link distance of $2$
km. For weak, moderate and strong FSO turbulence, channel
parameters are taken as $\sigma_x = 0.3$, $(\alpha,\beta) =
(4.16,2.21)$ and $E[I] = 1$, respectively. For RF environment, CM1
is assumed.} \label{sim_2}\end{centering}
\end{figure}

In Fig. \ref{sim_2}, the transmission of the detected optical
signals through a LOS RF UWB channel in the last-step to the
end-user is considered. The parameters for LOS channel model, CM1
are taken as $\sigma_g = 3$ dB, $A_0=47$ dB and $\psi = 1.7$
\cite{tarokh,tarokh2} and the RF link distance is taken as 2
meters. For the FSO link, we assume that the signals are
transmitted over 2 km at 30 and 45 dB. Under weak and moderate
conditions, $C_n^2$ is again chosen as $5.1 \times 10^{-15}$ and
$1.76 \times 10^{-14}$ which corresponds to $\sigma_x = 0.3$ and
$(\alpha,\beta) = (4.16,2.21)$, respectively. The results are
shown in Fig. \ref{sim_2} where the thick dashed curve shows the
performance of the RF UWB transmission over CM1 channels only. The
rest of the curves consider the case where the detected bit values
are transmitted over the FSO channel. Therefore, these curves
represent the combined error performance over both the FSO and the
RF sections of the transmission link. Notice that the hybrid
system performance exhibits error floors resulting from the errors
introduced in FSO links. However, if the SNR over the FSO link is
over $30$ dB, the error floor for weak turbulence is significantly
low. This indicates at this or higher SNR, under weak turbulence,
the hybrid system increases the UWB signalling range to 2 km
without any sacrifice in the performance. Similar results can be
achieved for moderate turbulence conditions, if the FSO channel
SNR is $45$ dB or more. Unfortunately, for the extreme case of
strong turbulence, the error floors appear high even at large SNR
values.

\begin{figure}[ht!]
\begin{centering}
\includegraphics[height=3.5in]{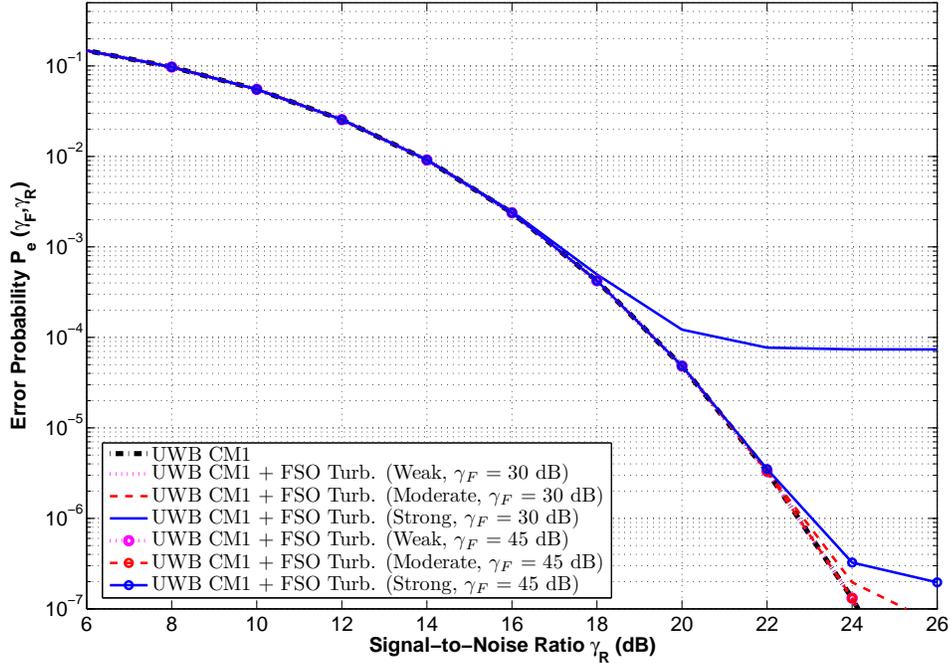}
\caption{$(27,31)_8$ convolutional coded FSO + RF UWB system
performance over 2 km.} \label{figcoded}\end{centering}
\end{figure}

To lower these error floors, we consider the case when  a simple
convolutional encoder/Viterbi decoder pair within the FSO
subsystem is employed. With this addition, the errors introduced
in the FSO links are reduced to improve the overall system
performance. A rate $1/2$ convolutional code with generator
$(27,31)_8$ and $N_b = 2$, $d_{free} = 7$ (\cite{costello}, pp.
540) is employed together with a Viterbi decoder. The coded system
performance under each turbulence regime is shown in Fig.
\ref{figcoded} and only the error floors under strong turbulence
conditions are shown for comparison purposes with uncoded
signalling scheme. Notice that the use of even a very simple code
reduces the error floors significantly. For instance, the error
floors for the strong turbulence condition at $\gamma_F = 30$ dB
is $1.5 \times 10^{-2}$ for the uncoded system whereas it is
reduced significantly down to $7.4 \times 10^{-5}$ when coding is
employed. Table \ref{errorfloors} shows the error floors in both
cases for each turbulence condition. Thus, at the cost of
additional complexity introduced by Viterbi decoder, the error
floors of the uncoded system performance is significantly lowered.

\begin{table}[ht!]
    \addtolength\belowcaptionskip{8pt}
    \centering \caption{Error floors of the composite system error
    performance for the uncoded and coded cases.}
     \begin{tabular}{c | c | c| c| c}
        \hline\hline   &  & Weak & Moderate & Strong
        \\ [0.5ex] \hline
        & Uncoded &  $7.4 \times 10^{-7}$ & $1.8 \times 10^{-3}$ & $1.3 \times 10^{-2}$ \\
        [-1ex] \raisebox{1.5ex}{$\gamma = 30$ dB} & Coded & $9.0 \times
        10^{-20}$& $6.5 \times 10^{-8}$ & $7.4 \times 10^{-5}$ \\[1ex]
        \hline
        & Uncoded &  $ 1.7 \times 10^{-13}$ & $5.0 \times 10^{-5}$   & $2.5\times 10^{-3}$ \\
        [-1ex] \raisebox{1.5ex}{$\gamma = 45$ dB} & Coded &
        $4.8 \times 10^{-43}$& $2.2 \times 10^{-13}$ & $2.0 \times 10^{-7}$ \\
        \hline \hline
    \end{tabular}
    \label{errorfloors}
\end{table}

\section{Channel Capacity Analysis}
In his pioneering work in 1948, Claude Shannon introduced the term
``channel capacity" where the bandwidth and the signal-to-noise
ratio together determine the quality of a transmission channel
\cite{shannon}. Until that moment, this relation was not seen. In
his works, the capacity of a noisy channel is defined as the
maximum possible transmission rate at which reliable communication
over the channel is possible \cite{proakis}. For transmission
rates that are lower than the channel capacity $C$, one can always
design codes that achieve arbitrarily small error probabilities.
Hence, for a discrete memoryless channel, the average channel
capacity can be expressed as \bea C = \max_{p(x)} I(X;Y) \eea
where $I(X;Y)$ denotes the mutual information between the random
variables $X$ and $Y$ which are commonly referred as the channel
inputs and channel outputs, respectively. The probability
distribution function denoted by $p(x)$ represents the channel
variations. The average capacity of a power-limited AWGN channel
is given by \bea C = \frac{1}{2} \log \left(1+\frac{P}{N} \right)
\eea where $P$ and $N$ denote the power of the signal and noise,
respectively and $P/N$ shows the signal-to-noise ratio (SNR). In
this form, the capacity is in bits$/$sec$/$Hz units if the
logarithm is taken in base 2 and in nats$/$sec$/$Hz units if the
natural logarithm is used. For a given bandwidth $B$, the same
channel capacity can be represented as \bea C = B \log
\left(1+\frac{P}{N} \right)\eea and in this form, its is in
bits$/$sec units since the bandwidth is used in the equation. For
different channel variations, the above can be revised to include
the effects of channel that are represented by $h$ as \bea E
\left[ C \right] = \int B \log\left(1 + h \frac{P}{N} \right)p(h)
\mathrm{d}h\eea This means that the average capacity of a fading
channel is found by taking the expectation of the AWGN channel
capacity over the fading term. This is reasonable to make a
comparison between the channel capacity of any fading channel and
AWGN channel. The former is always greater or equal to the latter
capacity and to prove this Jensen's inequality can be used such as
\bea E \left[ C \right] = \int B \log\left(1 + h \frac{P}{N}
\right)p(h) \mathrm{d}h \leq B \log\left(1 + \frac{P}{N} \int h
\mathrm{d}h \right) = B \log \left( 1 + \frac{P}{N} \right) \eea
and using the above equation, it can be concluded that the average
capacity of an AWGN channel is an upper bound to any fading
channel capacity. In the following subsections, the average
capacity of the proposed free-space optical link is investigated
under weak and moderate turbulence conditions.

\subsection{Channel Capacity Under Weak Turbulence Conditions}
On the propagation path of the transmitted light in FSO channel,
there are several inhomogeneities in the atmosphere which
independently lead to distributed phase delay and scattering.
Using this and invoking the Central Limit Theorem (CLT) then $X$
is Gaussian distributed random variable with mean $\mu_x$ and
variance $\sigma_X^2$. These inhomogeneities in the atmosphere are
related to the received light intensity by
\begin{align} I = I_0 \exp(2 X). \end{align} Here, it needs to be mentioned that in previous sections
the above equation included the mean term where in this section it
is ignored to include its effects in the closed-form expression of
the channel capacity. The marginal distribution of the light
intensity can be represented as
\begin{align} f_I(I) = \frac{1}{2I\sqrt{2\pi
\sigma_x^2}}\exp\left(-\frac{(\ln(I/I_0)-2\mu_X)^2}{8\sigma_x^2}\right).
\end{align}
Taking the expectation over the pdf of received light intensity
would result in
\begin{align} E[I] = \int_0^{\infty} \frac{I}{2I\sqrt{2\pi
\sigma_x^2}}\exp\biggl(-\frac{(\ln(I/I_0)-2\mu_X)^2}{8\sigma_x^2}\biggr)
\mathrm{d}I. \end{align} Applying change of variables such that
$y=\ln(I/I_0)-2\mu_x$ then $\mathrm{d}I = I_0
\exp\left(y+2\mu_x\right)\mathrm{d}y$. Hence,
\begin{align} E[I] = \int_{-\infty}^{\infty} \frac{I_0 e^{2\mu_x}}{2\sqrt{2\pi
\sigma_x^2}}\exp\biggl(-\frac{y^2-8 y \sigma_X^2 }{8\sigma_x^2
}\biggr) \mathrm{d}y. \end{align} Completing the squares in the
exponential term would result in
\begin{align} E[I] &= I_0 e^{2\mu_x} \int_{-\infty}^{\infty} \frac{1}{2\sqrt{2\pi
\sigma_x^2}}\exp\biggl(-\frac{(y-4\sigma_X^2)^2}
{8\sigma_x^2}\biggr) e^{2\sigma_X^2}\mathrm{d}y \nonumber\\ &= I_0
e^{2\mu_x+2\sigma_X^2}\int_{-\infty}^{\infty}f_Y(y) \mathrm{d}y
\nonumber\\ E[I] &= I_0 e^{2\mu_x+2\sigma_X^2}.
\end{align}
Consider the case when the fading does not attenuate or amplify
the average power, then normalizing the fading coefficient such
that $E[I] = I_0$ would require to choose the mean of the
log-amplitudes to be $\mu_X = -\sigma_X^2$ \cite{uysal}. Under
this condition, the average capacity of the log-normal channel can
be calculated as \begin{align}E\left[ C \right] = \int_0^{\infty}
\frac{B \log_2 \left(1 + N_s \gamma_p \left(\eta I \right)^2
\right)}{2I\sqrt{2\pi
\sigma_x^2}}\exp\biggl(-\frac{(\ln(I/I_0)+2\sigma_X^2)^2}{8\sigma_x^2}\biggr)\mathrm{d}I.
\label{cap_lognormal}\end{align} The above expresses the capacity
in bits$/$sec units since the logarithm is taken in base $2$ and
the bandwidth of the message $B$ is considered in the equation.
Applying change of variables such that $y = \left(\ln (I/I_0)
+2\sigma_X^2\right)/\sqrt{8\sigma_x^2}$, then
(\ref{cap_lognormal}) becomes \begin{align} E\left[ C \right] =
\int_{-\infty}^{\infty} \frac{B}{\sqrt{\pi}} \log_2 \left(1 + N_s
\gamma_p \left(\eta I_0 e^{y \sqrt{8\sigma_x^2}-2\sigma_X^2}
\right)^2 \right)\exp\left(- y^2 \right)\mathrm{d}y
\label{cap_lognormal2}.\end{align} Integrals of this form can be
approximated via Gauss-Hermite expression form as in (\ref{herm}).
Then, rewriting (\ref{cap_lognormal2}) will result in
\begin{eqnarray}
E\left[ C \right] = B \sum_{i = 1}^{n} \frac{\omega_i}{\sqrt{\pi}}
\hspace{.1cm}  \log_2 \left(1 + N_s \gamma_p \left(\eta I_0
e^{y_i\sqrt{8\sigma_x^2}-2\sigma_X^2} \right)^2
\right)\label{cap_lognormal_hermite}
\end{eqnarray}
where $y_i$ and $\omega_i$ are as described in previous section.
\subsection{Channel Capacity Under Moderate Turbulence Conditions}
The channel capacity of an FSO link under moderate turbulence
conditions is calculated by solving \begin{align} E[C] = B \times
\frac{2(\alpha
\beta)^{\frac{\alpha+\beta}{2}}}{\Gamma(\alpha)\Gamma(\beta)\ln(2)}
\int_0^{\infty} I^{\frac{\alpha+\beta}{2}-1} \ln\left(1+ N_s
\gamma_F (\eta I)^2\right) K_{\alpha-\beta}(2\sqrt{\alpha \beta
I}) \mathrm{d}I.
\end{align}
where the pdf of the light intensity under moderate conditions is
used in the integration. For the above equation, applying change
of random variables such that $V = I^2$ and Meijer's G-function
representations of the integrands $\ln(\cdot)$ and $K_n(\cdot)$,
it can be written as
\begin{align}\label{integral_capacity_1} E[C] = B \times
\frac{(\alpha
\beta)^{\frac{\alpha+\beta}{2}}}{\Gamma(\alpha)\Gamma(\beta)\ln(2)}
& \int_0^{\infty} V^{\frac{\alpha+\beta}{4}-1}
G_{2,2}^{1,2}\left(N_s \gamma_F \eta^2 V\bigg| \bary{l}1,1\\1,0
\eary \right)  \\ & \times \frac{1}{2} G_{0,2}^{2,0}\left(\alpha
\beta \sqrt{V} \bigg|\bary{l}-\\
\frac{\alpha-\beta}{2},\frac{{\beta-\alpha}}{2} \eary \right)
\mathrm{d}V. \nonumber
\end{align}
This form can be further simplified to a closed-from expression
such as
\begin{align}\label{integral_capacity_3} E[C] = &
\frac{B}{4 \pi \Gamma(\alpha)\Gamma(\beta)\ln(2)}
\left(\frac{\alpha \beta}{\sqrt{\eta^2 N_s
\gamma_F}}\right)^{\frac{\alpha+\beta}{2}} \\ & \times
G^{6,1}_{2,6}\left(\frac{(\alpha \beta)^2}{16 N_s \gamma_F \eta^2}
\bigg|
\bary{l}\frac{-\alpha-\beta}{4},\frac{4-\alpha-\beta}{4}\\\frac{\alpha-\beta}{4},\frac{\alpha-\beta+2}{4},
\frac{\beta-\alpha}{4},\frac{\beta-\alpha+2}{4},\frac{-\alpha-\beta}{4},\frac{-\alpha-\beta}{4}
\eary \right). \nonumber
\end{align}
where the intermediate steps to derive the above closed-form
expression are shown in Appendix \ref{appendix channel capacity}
in detail.

\begin{figure}[h!]
\begin{centering}
\includegraphics[height = 4in]{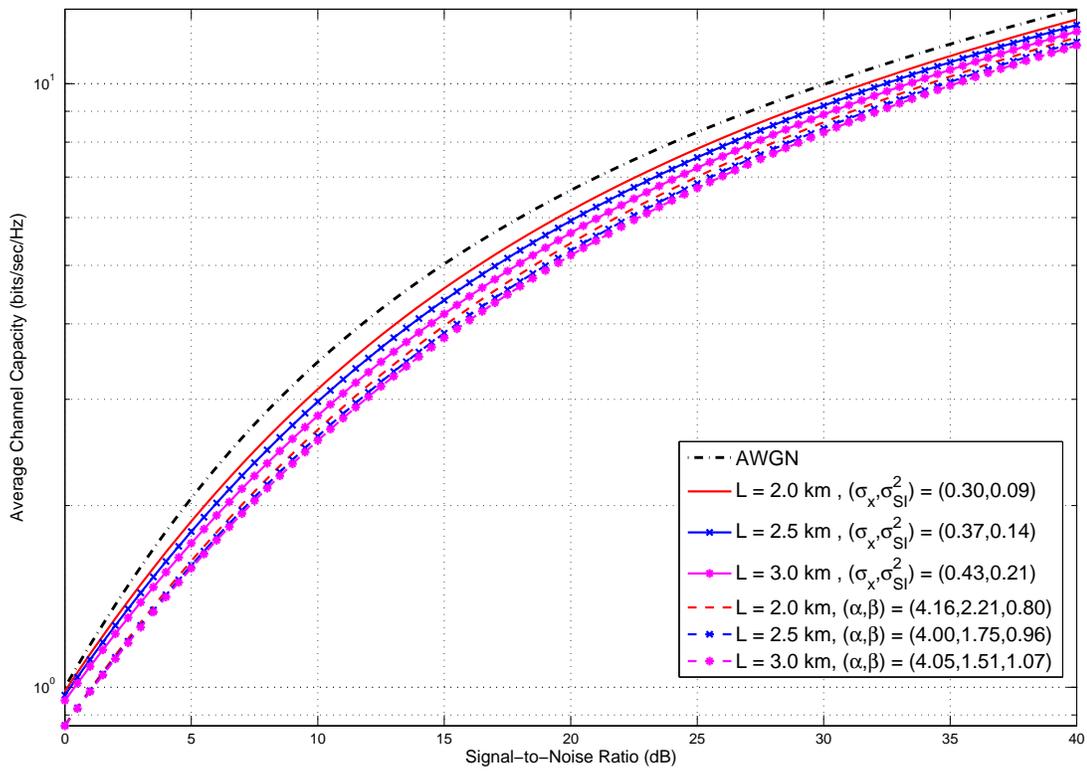}
\caption{Average channel capacity under weak and moderate
turbulence conditions for different link distances.}
\label{channel capacity Weak and Moderate}
\end{centering}
\end{figure}

In Figure \ref{channel capacity Weak and Moderate}, the average
channel capacity of the FSO link is presented. Depending on the
turbulence conditions, different fading effects are investigated.
Under weak and moderate turbulence conditions, the average channel
capacity of the FSO system are plotted. The wavelength is taken as
$1550$ nm and the wavenumber spectrum structure parameters for
weak and moderate turbulence are taken as $C_n^2 = 5.1 \times
10^{-15}$ and $C_n^2 = 1.76 \times 10^{-14}$, respectively. The
link distances of $2, 2.5$ and $3$ are considered in the figure.
As expected, when the system is under weak turbulence condition,
the average channel capacity is higher compared to the case under
moderate conditions. This behavior is clearly depicted in the
figure. Also, another way to interpret the figure is the increase
in the link distance results in increase in the scintillation
index and this decreases the channel capacity. This is due to the
fact that scintillation effects causing amplitude variations
become more dominant to reduce the channel capacity as the
scintillation index increases. The AWGN channel capacity is also
included in the figure to show that it is the upper bound and
small scintillation index values yield close average capacities to
this bound.

%
%
%

\chapter{CONCLUSIONS}\label{section conclusion}
This thesis outlines the benefits of UWB systems that operate in
the unlicensed spectrum which can provide very high data speeds
and along with the great interest from both the research community
and the industry, UWB systems clearly have a potential in various
fields of electrical engineering such as WPANs, sensor networks
including tracking and ranging applications, vehicular radar
systems and imaging systems. However, UWB systems that operate in
RF channels are limited to short distances. To this end, a novel
optical UWB system model that uses FSO signals is presented and
the proposed approach intends to extend the link distance from
couple of meters to several kilometers. Compared to RF links, FSO
systems are not subject to severe multipath interference that
inhibits the transmission of pulses over long ranges but on the
other hand, they suffer from atmospheric turbulence fading that
severely degrade the link performance.

Along with the necessary channel parameter definitions and their
mathematical expressions, statistical distributions to model
different turbulence conditions in FSO channels are also included
in this thesis. Using these models, the error probability analysis
of the proposed FSO system model under weak, moderate and strong
turbulence induced fading conditions are presented. Depending on
the system design, an extended scenario where the estimates of the
FSO link outputs are further distributed over RF links to the
end-user in the last-step is also considered and its error
analysis is also included. Both the theoretical and simulation
results of two scenarios depict that increase in the scintillation
index which indicates the strongness of the turbulence effects
closely influence the system performance. As the scintillation
index increases, bit error probability of the FSO system
significantly increases and saturates for the strong turbulence
conditions. Also, under weak turbulence regime, the effect of the
errors introduced in FSO link are almost negligible but as the the
turbulence conditions get worse, high error floors are observed in
the system performance. A simple convolutional encoder and Viterbi
decoder pair in FSO link is proposed as a solution to protect the
bits and reduce the error floors. When error-control coding is
employed, it is observed that the error floors are significantly
lowered and the quality of service of the system is improved. It
should be emphasized that better error correcting codes such as
LDPC coding can further decrease the error probability at the cost
of increased complexity.

Another important system parameter, the average channel capacity
of the proposed FSO system is investigated and its closed form
expressions under weak and moderate turbulence conditions are also
presented in this thesis. Concluding remarks of this analysis are
that the average channel capacity without any fading is the upper
bound to the fading channels and increase in the scintillation
index gradually decreases the FSO channel capacity. Although there
are some recent studies on analyzing the average channel capacity
of FSO links under weak and moderate turbulence conditions such as
in \cite{uysal_capacity,gamma_nista}, the outage channel capacity
analysis still remains unanswered. Also, the average channel
capacity of the composite system consisting of FSO and RF links is
not investigated yet. Several studies propose hybrid link
structure that use RF links to back up whenever FSO links are not
available. These studies focus on narrowband transmission of
signals that are not applicable to UWB signalling. The link
availability drawback of FSO systems arise as a consequence of
weather conditions on the chosen site. Via employing several FSO
transceivers in a mesh topology, link availability problem can be
reduced.

The studies on FSO systems will clearly increase the knowledge on
atmospheric effects and the physics of turbulence regimes, improve
better techniques to generate optical pulse that advance the
technology of FSO devices. It is obvious that FSO system
performance will continue to improve reaching higher data rates,
better accuracy in aligning devices and more importantly link
availability under all weather conditions.

\appendix      

\chapter{Mellin Transform and Applications} The
properties of Mellin transform are very similar to the ones of
Laplace and Henkel transforms. The Mellin transform of the
function $f(x)$ and the inverse Mellin transform of $F(s)$ are
defined as \cite{andrews_math_tech}
\begin{subequations}\begin{align} M\{f(x)\} &= F(s) =
\int_0^{\infty} x^{s-1}f(x)\mathrm{d}x \\ M^{-1}\{F(s)\} &=
f(x)U(x) = \frac{1}{2\pi i} \int_{c-i
\infty}^{c+i\infty}x^{-s}F(s) \mathrm{d}s
\end{align}\end{subequations}
Some basic properties of Mellin transform are listed in the below
table.
\begin{table}[h!]\label{Melintrasformtable}\caption{Basic Properties of the Mellin Transform}\begin{center}
\begin{tabular}{|l|l|}\hline
Linearity & $M\{C_1 f(x) + C_2 g(x) \} = C_1 F(s) + C_2 G(s)$
\\\hline
Scaling & $ M\{f(ax)\} = (1/a^s) F(s)$, $a>0$ \\\hline Translation
& $M\{x^a f(x) \} = F(s+a)$,  $a>0$ \\\hline Derivative &
$M\{f^{(n)}(x)\} = (-1)^n \frac{\Gamma(s)}{\Gamma(s-n)}F(s-n)$, $n
= 1,2,3,\ldots$
\\\hline
Convolution & $M\{\int_0^{\infty}
f(\frac{x}{u})g(u)\frac{\mathrm{d}u}{u} \} = F(s)G(s)$
\\\hline
\end{tabular}\end{center}\end{table}
The following table shows some special pairs that are useful while
working the Mellin transform.
\begin{table}[h!]
\caption{Special Mellin Transform Pairs}\begin{center}
\begin{tabular}{|l|l|}\hline \label{Melinspecialtrasformtable}
$F(s) = M\{f(x) \}$ & $f(x) = M^{-1}\{F(s)\}$ \\\hline
$\pi/\sin(\pi s)$ & $1/(1+x)$ \\\hline $\Gamma(s)/a^s$, $a>0$ &
$e^{-ax}$ \\\hline $\Gamma(s) \cos(\pi s/2) / a^s$, $a>0$ &
$\cos(ax)$ \\\hline $\Gamma(s) \sin(\pi s/2) / a^s$, $a>0$ &
$\sin(ax)$ \\\hline $-\frac{\sqrt{\pi}}{8}
\frac{\Gamma(s/4)}{\Gamma(1/2-s/4)}$ & $\sin^2(x^2)$ \\\hline
\end{tabular}\end{center}\end{table}
\section{Electromagnetic Wave Propagation}\label{Mellin transfomr in wave propagation}
In a random medium, modelling wave propagation leads to difficult
integrals that cannot be easy solved by classical methods. Even
numerical approaches may be hard to apply because in most cases,
the integrand is defined as the difference between two quantities
that each may produce a divergent integral or sometimes, the
integrand may be  a product of two functions which one goes to
infinity and the other goes to zero. For such problems, Mellin
transform may provides an easy method. These problems commonly
arise in wave propagation. The Rytov variance of random
fluctuations can be found by solving the below integral
\cite{andrews_special}\bea
\label{mellin_integral_1}\sigma_I^2 = 8 \pi^2 k^2 L \int_0^1
\int_0^{\infty} \kappa \Phi_n(\kappa) \left( 1 - \cos\left(\frac{L
\kappa^2 \xi}{k} \right) \right)\mathrm{d}\kappa \mathrm{d}\xi\eea
where $k$ is the optical wave number, $L$ is the propagation path
length, $\xi$ is the normalized distance variable and
$\Phi_n(\kappa)$ is the spatial spectrum of refractive-index
fluctuations. Using Kolmogorov power-law spectrum,
$\Phi_n(\kappa)$ can be expressed as \bea
\label{mellin_kolmogorov} \Phi_n(\kappa) = 0.033 C_n^2
\kappa^{-11/3} \eea where $C_n^2$ is the refractive-index
structure parameter and it is a measure of the strength of random
fluctuations of the medium. Using the trigonometric identity such
that $2 \sin^2(x/2) = 1 - \cos(x)$ in (\ref{mellin_integral_1})
and employing (\ref{mellin_kolmogorov}), the below integral
represent the variance of random fluctuations such as
\begin{subequations}\begin{align} \sigma_I^2 &= 16 \pi^2 k^2 L (0.033) C_n^2
\int_0^1 \int_0^{\infty} \kappa^{-8/3} \sin^2\left(\frac{L\kappa^2
\xi}{2 k} \right) \mathrm{d}\kappa \mathrm{d}\xi
 \\& = 5.211 C_n^2 k^2 L \left(\frac{L}{2k}\right)^{5/6}\int_0^1 \xi^{5/6}
 \mathrm{d}\xi \int_0^\infty x^{-8/3} \sin^2(x^2)\mathrm{d}x \end{align}\end{subequations}
where change of random variables method is applied to the second
step. The first integral is trivial \bea \int_0^1 \xi^{5/6}
 \mathrm{d}\xi = \frac{6}{11}\eea  For the second integral, the
Mellin transform can be directly applied via Table
\ref{Melinspecialtrasformtable} such that
 \begin{subequations}\begin{align}
 \int_0^\infty x^{-8/3} \sin^2(x^2)\mathrm{d}x &=
M\left[\sin^2(x^2); s=-\frac{5}{3}\right] \\ &= - \lim_{s
\rightarrow -\frac{5}{3}}
\frac{\sqrt{\pi}}{8}\frac{\Gamma(s/4)}{\Gamma(1/2-s/4)}.
\end{align}\end{subequations}
Combining above equations, the Rytov variance for plane waves can
be represented in its final form as \bea \sigma_I^2 = 1.23 C_n^2
k^{7/6} L^{11/6}. \eea
\chapter{Meijer's G-Function and Applications}
The Meijer's G-function was defined by the Dutch mathematician
Cornelis Simon Meijer in 1936 to introduce a general function that
includes most of the known special functions as particular cases.
His definition was able to include many known special functions as
particular cases. This first definition which was using series
expansion was made revised by Arthur Erdélyi in 1953 to take
integrals in the complex plane. With this new current definition,
most of the special functions can be expressed in terms of the
G-function and of the Gamma function. The Meijer's G-function in
path integral form can be expressed as \bea G^{m,n}_{p,q}\left(z
\bigg|\bary{l}a_1,\ldots,a_p\\b_1,\ldots,b_p\eary \right) =
\frac{1}{2\pi i} \int_L \frac{\prod_{j=1}^m
\Gamma(b_j-s)\prod_{j=1}^n\Gamma(1-a_j+s)}{\prod_{j=m+1}^q
\Gamma(1-b_j+s)\prod_{j=n+1}^p \Gamma(a_j-s)}z^s \mathrm{d}s \eea
where $L$ denotes the integration path, $\prod$ is the product
operator for sequences and $\Gamma(\cdot)$ is the gamma function.
The above definition is also referred to as the Mellin transform
and it holds under the following three assumptions.
\begin{enumerate}\item m,n,p and q are all integers (ie, m,n,p,q $\in Z$) and $0<m<q$ and $0<n<p$,
\item No pole of any $\Gamma(b_j-s)$ can coincide with any pole of
$\Gamma(1-a_k-s)$, which means that $a_k-b_j \neq 1,2,3$ for
$k=1,2,\ldots,n$, \item $z \neq 0$.
\end{enumerate}
The integral path L has three possible choices are such as
\begin{enumerate}\item L goes from $-i\infty$ to $i \infty$ such as all
poles of $\Gamma(b_j-s)$ are on the right of the path, while the
poles of $\Gamma(1-a_k+s)$ are on the left for $j= 1,2,\dots,m$
and $k=1,2,\ldots,n$. For this path, the integral converges when
$|\arg z| < \delta \pi$ where \bea \label{delta} \delta = m + n -
\frac{1}{2}(p+q) > 0\eea The integral also converges when $| \arg
z| = \delta \pi \geq 0$ if \bea (q-p)\left(\sigma +
\frac{1}{2}\right)
> Re\{\upsilon\} +1 , \eea and $\sigma$ is the real part of integration variable,
$Re\{s\}$ and $s$ approaches to both $+i\infty$ and $-i\infty$
where \bea \label{upsilon} \upsilon = \sum_{j=1}^{q} b_j -
\sum_{j=1}^{p}a_j \eea

\item L begins and ends at $+\infty$ and encircles all poles of
$\Gamma(b_j-s)$ in negative direction exactly ones but does not
encircle the poles of $\Gamma(1-a_k+s)$ for $j= 1,2,\dots,m$ and
$k=1,2,\ldots,n$. Now, the integral path converges for all values
of z, if $q>p\geq0$. Another convergent path is when $q=p>0$ for
$|z|<1$. Also, the path that $|z|=1$ converges for
$Re\{\upsilon\}<-1$ where $p=q$ and $\upsilon$ is as defined above
in (\ref{upsilon}).

\item L begins and ends at $-\infty$ and encircles all poles of
$\Gamma(1-a_k+s)$ in positive direction exactly ones but does not
encircle the poles of $\Gamma(b_j-s)$ for $j= 1,2,\dots,m$ and
$k=1,2,\ldots,n$. Now, the integral path converges for all values
of z, if $p>q\geq0$. Another convergent path is when $q=p>0$ for
$|z|>1$. Also, the path that $|z|=1$ converges for
$Re\{\upsilon\}<-1$ where $p=q$ and $\upsilon$ is as defined above
in (\ref{upsilon}).
\end{enumerate}
There are several special functions that can be expressed in terms
of Meijer's G-function such as \begin{align} e^z &=
G_{0,1}^{1,0}\left(-z\bigg| \bary{l} - \\ 0 \eary \right)\\
\frac{z^b}{\sqrt{\pi}}\cos(2\sqrt{z}) &=
G^{1,0}_{0,2}\left(z\bigg|\bary{l}-\\b,b+\frac{1}{2}\eary \right)\\
\frac{z^{b-\frac{1}{2}}}{\sqrt{\pi}}\sin(2\sqrt{z}) &=
G^{1,0}_{0,2}\left(z\bigg|\bary{l}-\\b,b-\frac{1}{2}\eary \right)\\
\ln(1+z) &= G^{1,2}_{2,2} \left(z \bigg| \bary{l}1,1\\1,0 \eary
\right) \label{meijer_ln()}\\
K_\upsilon(z) &= \frac{1}{2} G^{2,0}_{0,2} \left(\frac{z^2}{4}
\bigg| \bary{l}-\\ \frac{\upsilon}{2},\frac{-\upsilon}{2} \eary
\right) \label{meijer_Kn}\end{align} The following identities are
useful to adjust the powers and arguments of Meijer's G-function
such as
\begin{align} z^{\alpha} G^{m,n}_{p,q}\left(z\bigg|\bary{l}
\overrightarrow{a_p}
\\ \overrightarrow{b_q} \eary\right) &=
G^{m,n}_{p,q}\left(z\bigg|\bary{l} \overrightarrow{a_p}+\alpha \\
\overrightarrow{b_q}+\alpha \eary\right) \label{z^a} \\
G^{m,n}_{p,q}\left( z\bigg|\bary{l} \overrightarrow{a_p}
\\ \overrightarrow{b_q} \eary\right) &=
\frac{k^{1+\upsilon+(p-q)/2}}{(2\pi)^{(k-1)\delta}} \times \label{z^k} \\
\nonumber &
G^{km,kn}_{kp,kq}\left(\frac{z^k}{k^{k(q-p)}}\bigg|\bary{l}
\frac{a_1}{k},\ldots,\frac{a_1+k-1}{k},\ldots,\frac{a_p}{k},\ldots,\frac{a_p-k+1}{k}
\\ \frac{b_1}{k},\ldots,\frac{b_1+k-1}{k},\ldots,\frac{b_q}{k},\ldots,\frac{b_q-k+1}{k}
\eary\right) \end{align} where $\alpha$ in (\ref{z^a}) is a
constant and
 $\delta$ and $\upsilon$ in (\ref{z^k}) are as defined
previously in (\ref{delta}) and (\ref{upsilon}). The following
equation is often used while taking the definite integral of the
product of any two Meijer's G-function. Via convolution theorem,
the integral of the product of any two Meijer's G-function can be
represented in a single Meijer's G-function in its closed form
such as \begin{align}\label{integralWiki} \int_0^{\infty}
&G^{m,n}_{p,q}\left( \eta^* z \bigg|\bary{l} \overrightarrow{a_p}
\\ \overrightarrow{b_q}\eary\right)  G^{\mu,\nu}_{\sigma,\tau}\left(\omega z\bigg|\bary{l} \overrightarrow{c_\sigma}
\\ \overrightarrow{d_\tau} \eary\right) \mathrm{d}z  \\ &=
\frac{1}{\eta^*} G^{n+\mu,m+\nu}_{q+\sigma,p+\tau}\left(
\frac{\omega}{\eta^*} \bigg|\bary{l}
-b_1,\ldots,-b_m,\overrightarrow{c_\sigma},-b_{m+1},\ldots,-b_q
\\ -a_1,\ldots,-a_n,\overrightarrow{d_\tau},-a_{n+1},\ldots,-a_p \eary\right)\nonumber\\
&= \frac{1}{\omega} G^{m+\nu,n+\mu}_{p+\tau,q+\sigma}\left(
\frac{\eta^*}{\omega} \bigg|\bary{l}
a_1,\ldots,a_n,-\overrightarrow{d_\tau},a_{n+1},\ldots,a_p
\\ b_1,\ldots,b_m,-\overrightarrow{c_\sigma},b_{m+1},\ldots,b_q
\eary\right)\nonumber \end{align} Above two equations can be
written in a more compact way using (\ref{z^a}) and
(\ref{integralWiki}) as
\begin{align}\label{integralWolf} \int_0^{\infty}
& z^{f-1}G^{m,n}_{p,q}\left( \eta^* z \bigg|\bary{l}
\overrightarrow{a_p}
\\ \overrightarrow{b_q}\eary\right)  G^{\mu,\nu}_{\sigma,\tau}\left(\omega z\bigg|\bary{l} \overrightarrow{c_\sigma}
\\ \overrightarrow{d_\tau} \eary\right) \mathrm{d}z  \\ &=
\eta^{* -f} G^{n+\mu,m+\nu}_{q+\sigma,p+\tau}\left(
\frac{\omega}{\eta^*} \bigg|\bary{l} c_1,\ldots,c_\nu, 1-f-b_1,
\ldots, 1-f-b_q,c_{\nu+1},\ldots,c_\sigma \\ d_1,\ldots, d_\mu,
1-f-a_1, \ldots, 1-f-a_p, d_{\mu+1},\ldots,d_\tau
\eary\right)\nonumber\end{align} This integral form is quite
useful to derive a closed form solution for the error probability
for the free space optical links under moderate turbulence
conditions. In the following section, detailed derivation for this
condition is presented.
\newpage
\section{Error Probability Analysis Under Moderate Turbulence}\label{appendix error probability}
The integrands $\erfc(\cdot)$ and $K_n(\cdot)$ are represented in
terms of Meijer's G-function, the error probability of FSO link
under moderate conditions becomes {\begin{align}
P_{e,F}(\gamma_{F})&=\frac{2(\alpha\beta)^{\frac{\alpha+\beta}{2}}}{\Gamma(\alpha)\Gamma(\beta)}
\int_0^\infty I^{\frac{\alpha+\beta}{2}-1}
\frac{1}{2}G^{2,0}_{0,2}\bigg[\alpha\beta I\bigg|\begin{array}{c}
-\\\frac{\alpha-\beta}{2},\frac{\beta-\alpha}{2}
\end{array}\bigg]
\frac{1}{2\sqrt{\pi}}G^{2,0}_{1,2}\bigg[\frac{N_s \gamma_{F} (\eta
I)^2}{2}\bigg|\begin{array}{c}1\\0,\frac{1}{2}
\end{array}\bigg] \mathrm{d}I
\end{align}
where the first Meijer's G-function represents $K_n(\cdot)$ and
the second one stands for $\erfc(\cdot)$. In this form, the two
terms that are related to intensity, $I$ have different powers and
it can not be directly used in (\ref{integralWolf}). Let us
represent the term associated to $\erfc(\cdot)$ similar to the
form in (\ref{z^k}) as
\begin{align} G^{2,0}_{0,2}\bigg[\alpha\beta I\bigg|\begin{array}{c}
-\\\frac{\alpha-\beta}{2},\frac{\beta-\alpha}{2}
\end{array}\bigg] = \frac{1}{2\pi} \hspace{.05in} G^{4,0}_{0,4}\bigg[\frac{(\alpha\beta I)^2}{2^4} \bigg|\begin{array}{c}
-\\\frac{\alpha-\beta}{4},\frac{\alpha-\beta+2}{4},\frac{\beta-\alpha}{4},\frac{\beta-\alpha+2}{4}
\end{array}\bigg].\end{align}
Using this expression of $\erfc(\cdot)$, let us rewrite the
integral and also adjust the integration variable such that it is
same as the one in Meijer term, {\begin{align}
\label{pe_gamma_gamma_appendix}
P_{e,F}(\gamma_{F})=\frac{2(\alpha\beta)^{\frac{\alpha+\beta}{2}}}{\Gamma(\alpha)\Gamma(\beta)}
\times \frac{1}{8 \pi^{3/2}} \int_0^\infty &
I^{2[(\frac{\alpha+\beta}{4}+\frac{1}{2})-1]} \hspace{.05in}
G^{4,0}_{0,4}\bigg[\frac{(\alpha\beta I)^2}{2^4}
\bigg|\begin{array}{c} -
\nonumber\\\frac{\alpha-\beta}{4},\frac{\alpha-\beta+2}{4},\frac{\beta-\alpha}{4},\frac{\beta-\alpha+2}{4}
\end{array}\bigg] \\ &
G^{2,0}_{1,2}\bigg[\frac{N_s \gamma_{F} (\eta
I)^2}{2}\bigg|\begin{array}{c}1\\0,\frac{1}{2}
\end{array}\bigg] \mathrm{d}I.
\end{align}
Let us apply change of variables such that $V = I^2$ and
$\mathrm{d}I = \frac{1}{2\sqrt{V}}\mathrm{d}V$. {\begin{align}
\label{pe_gamma_gamma_appendix_2}
P_{e,F}(\gamma_{F})=\frac{2(\alpha\beta)^{\frac{\alpha+\beta}{2}}}{\Gamma(\alpha)\Gamma(\beta)}
\times \frac{1}{8 \pi^{3/2}} \int_0^\infty &
\frac{V^{[(\frac{\alpha+\beta}{4}+\frac{1}{2})-1]}}{2\sqrt{V}}
\hspace{.05in} G^{4,0}_{0,4}\bigg[\frac{(\alpha\beta)^2 V}{2^4}
\bigg|\begin{array}{c} -
\nonumber\\\frac{\alpha-\beta}{4},\frac{\alpha-\beta+2}{4},\frac{\beta-\alpha}{4},\frac{\beta-\alpha+2}{4}
\end{array}\bigg] \\ &
G^{2,0}_{1,2}\bigg[\frac{N_s \gamma_{F} \eta^2
V}{2}\bigg|\begin{array}{c}1\\0,\frac{1}{2}
\end{array}\bigg] \mathrm{d}V.
\end{align}
For the integral in this form, closed form expression in
(\ref{integralWolf}) can be directly applied. The power for the
square of intensity $I^2$ is the term represented by $f$ in
(\ref{integralWolf}) and for this integral, it is $f =
(\alpha+\beta)/4$, and the roots of the first Meijer's G-function
integrand are $R =
\{\frac{\alpha-\beta}{4},\frac{\alpha-\beta+2}{4},\frac{\beta-\alpha}{4},\frac{\beta-\alpha+2}{4}\}$.
The new roots for the result of the above integral will be
determined by $1-f-R_i$, $i\in \{1,2,3,4\}$ and they can be
expressed as
\begin{subequations}\begin{align}1-f-R_1 &=
1-\left(\frac{\alpha+\beta}{4}\right)-\left(\frac{\alpha-\beta}{4}
\right)= \frac{2-\alpha}{2} \\ 1-f-R_2 &=
1-\left(\frac{\alpha+\beta}{4}\right)-\left(\frac{\alpha-\beta+2}{4}
\right)= \frac{1-\alpha}{2} \\ 1-f-R_3 &=
1-\left(\frac{\alpha+\beta}{4}\right)-\left(\frac{\beta-\alpha}{4}
\right)= \frac{2-\beta}{2} \\ 1-f-R_4 &=
1-\left(\frac{\alpha+\beta}{4}\right)-\left(\frac{\beta-\alpha+2}{4}
\right)= \frac{1-\beta}{2}. \end{align}\end{subequations} Also,
the arguments of the Meijer's G-function terms in
(\ref{integralWolf}) are $\eta^* = (\alpha \beta)^2/16$ and
$\omega = \eta^2 N_s \gamma_{F}/2$. The argument of the closed
form expression will be $\omega/\eta^* = (8\eta^2 N_s
\gamma_{F})/(\alpha\beta)^2$. Using the above roots and arguments
of Meijer's G-function, (\ref{pe_gamma_gamma_appendix}) can be
expressed in its closed form as
\begin{align}P_{e,F}(\gamma_{F}) &=
\frac{2^{\alpha+\beta-3}}{\Gamma(\alpha)\Gamma(\beta) \pi^{3/2}}
\hspace{.05in} G^{2,4}_{5,2}\bigg[\frac{8\eta^2 N_s
\gamma_{F}}{\alpha^2
\beta^2}\bigg|\begin{array}{c}\frac{1-\alpha}{2},\frac{2-\alpha}{2},\frac{1-\beta}{2},\frac{2-\beta}{2},1\\0,\frac{1}{2}\end{array}\bigg].
\end{align}
\newpage
\section{Channel Capacity Under Moderate Turbulence}\label{appendix channel capacity}
The channel capacity of a FSO link under moderate turbulence
conditions is determined by solving \begin{align} E[C] = B \times
\frac{2(\alpha
\beta)^{\frac{\alpha+\beta}{2}}}{\Gamma(\alpha)\Gamma(\beta)\ln(2)}
\int_0^{\infty} I^{\frac{\alpha+\beta}{2}-1} \ln\left(1+ N_s
\gamma_F (\eta I)^2\right) K_{\alpha-\beta}(2\sqrt{\alpha \beta
I}) \mathrm{d}I.
\end{align}
Applying change of random variables such that $I^2 = V$ and
$\mathrm{d}I = \frac{1}{2\sqrt{V}}\mathrm{d}V$, then the channel
capacity can be expressed as
\begin{align} E[C] = B \times
\frac{2(\alpha
\beta)^{\frac{\alpha+\beta}{2}}}{\Gamma(\alpha)\Gamma(\beta)\ln(2)}
\int_0^{\infty}
\frac{V^{\frac{\alpha+\beta}{4}+\frac{1}{2}-1}}{2\sqrt{V}}
\ln\left(1+ N_s \gamma_F \eta^2 V\right)
K_{\alpha-\beta}(2\sqrt{\alpha \beta \sqrt{V}}) \mathrm{d}V.
\end{align}
In this form, the Meijer's G-function representation of
$\ln(\cdot)$ and $K_n(\cdot)$ can be introduced via
(\ref{meijer_ln()}) and (\ref{meijer_Kn}) such that
\begin{align}\label{integral_capacity_1} E[C] = B \times
\frac{(\alpha
\beta)^{\frac{\alpha+\beta}{2}}}{\Gamma(\alpha)\Gamma(\beta)\ln(2)}
& \int_0^{\infty} V^{\frac{\alpha+\beta}{4}-1}
G_{2,2}^{1,2}\left(N_s \gamma_F \eta^2 V\bigg| \bary{l}1,1\\1,0
\eary \right)  \\ & \times \frac{1}{2} G_{0,2}^{2,0}\left(\alpha
\beta \sqrt{V} \bigg|\bary{l}-\\
\frac{\alpha-\beta}{2},\frac{{\beta-\alpha}}{2} \eary \right)
\mathrm{d}V. \nonumber
\end{align}
The arguments of the two Meijer's G-function has different powers
of the intensity. Therefore, let us use (\ref{z^k}) to change the
argument for $K_n(\cdot)$ such that
\begin{align}
G_{0,2}^{2,0}\left(\alpha \beta
\sqrt{V} \bigg|\bary{l}-\\
\frac{\alpha-\beta}{2},\frac{{\beta-\alpha}}{2} \eary \right) =
\frac{1}{2 \pi} G_{0,4}^{4,0}\left(\frac{(\alpha \beta)^2 V}{16} \bigg|\bary{l}-\\
\frac{\alpha-\beta}{2},\frac{\alpha-\beta+2}{2},\frac{{\beta-\alpha}}{2},\frac{{\beta-\alpha+2}}{2}
\eary \right).
\end{align}
This form of $K_n(\cdot)$ can be used in
(\ref{integral_capacity_1}) to derive a closed form expression
such as
\begin{align}\label{integral_capacity_2} E[C] = B \times
\frac{(\alpha
\beta)^{\frac{\alpha+\beta}{2}}}{\Gamma(\alpha)\Gamma(\beta)\ln(2)}
& \int_0^{\infty} V^{\frac{\alpha+\beta}{4}-1}
G_{2,2}^{1,2}\left(N_s \gamma_F \eta^2 V\bigg| \bary{l}1,1\\1,0
\eary \right)  \\ & \times \frac{1}{4\pi} G_{0,4}^{4,0}\left(\frac{(\alpha \beta)^2 V}{16} \bigg|\bary{l}-\\
\frac{\alpha-\beta}{2},\frac{\alpha-\beta+2}{2},\frac{{\beta-\alpha}}{2},\frac{{\beta-\alpha+2}}{2}
\eary \right) \mathrm{d}V. \nonumber
\end{align}
For the above integral, previously derived closed form integration
formula in (\ref{integralWolf}) can be directly used where its
roots are  \begin{subequations}\begin{align} R_1 &= 1 - f - b_1 =
1 - \frac{\alpha +
\beta}{4}  - 1 =  \frac{- \alpha - \beta}{4}  \\
R_2 &= 1 - f - b_2 = 1 - \frac{\alpha + \beta}{4}  - 0 = \frac{4 -
\alpha - \beta}{4} \\ R_3 &= 1 - f - a_1 = 1 - \frac{\alpha +
\beta}{4}  - 1 = \frac{- \alpha - \beta}{4}  \\
R_4 &= 1 - f - a_2 = 1 - \frac{\alpha + \beta}{4}  - 1 = \frac{-
\alpha - \beta}{4}.
\end{align}\end{subequations}
Using these roots, (\ref{integral_capacity_2}) can be rewritten in
its closed form as
\begin{align}\label{integral_capacity_3} E[C] = &
\frac{B}{4 \pi \Gamma(\alpha)\Gamma(\beta)\ln(2)}
\left(\frac{\alpha \beta}{\sqrt{\eta^2 N_s
\gamma_F}}\right)^{\frac{\alpha+\beta}{2}} \\ & \times
G^{6,1}_{2,6}\left(\frac{(\alpha \beta)^2}{16 N_s \gamma_F \eta^2}
\bigg|
\bary{l}\frac{-\alpha-\beta}{4},\frac{4-\alpha-\beta}{4}\\\frac{\alpha-\beta}{4},\frac{\alpha-\beta+2}{4},
\frac{\beta-\alpha}{4},\frac{\beta-\alpha+2}{4},\frac{-\alpha-\beta}{4},\frac{-\alpha-\beta}{4}
\eary \right). \nonumber
\end{align}

\end{document}